\documentclass[a4paper]{article}
\usepackage{geometry}
\usepackage{graphicx}
\usepackage{amsfonts,amsmath}
\usepackage{subfigure}

\usepackage{float}
\usepackage{hyperref}
\usepackage[table,dvipsnames]{xcolor}

\usepackage{xspace}
\usepackage{url}
\usepackage{color}

\usepackage{algorithmicx}
\usepackage{algorithm}
\usepackage{algpseudocode}

\newcommand{\ie}{i.\,e.\xspace}
\newcommand{\eg}{e.\,g.\xspace}
\newcommand{\etal}{et al.\xspace}
\newcommand{\nwk}{\textsf{NetworKit}\xspace}
\newcommand{\recon}{\textsf{ReCoN}\xspace}
\newcommand{\plm}{\textsf{PLM}\xspace}
\newcommand{\musk}{\textsf{MUSKETEER}\xspace}

\newcommand{\cc}{\textsf{c}}

\usepackage{authblk}

\title{Generating realistic scaled complex networks
\thanks{This work is partially supported by the DFG under grants ME~3619/3-1 and WA~654/22-1 within the Priority Program 1736 \emph{Algorithms for Big Data}, and by the NSF awards \#1522751 and \#1647361. Funding was also provided by \textit{Karlsruhe House of Young Scientists} via the \textit{International Collaboration Package}.
A preliminary version of the paper was presented at the 5th International Workshop on Complex Networks and their Applications~\cite{staudt2016generating}.
}
}

\author[2]{Christian L. Staudt}
\author[2]{Michael Hamann}
\author[1]{Alexander Gutfraind}
\author[3]{Ilya Safro}
\author[2]{Henning Meyerhenke}

\affil[1]{Laboratory for Mathematical Analysis of Complexity and Conflicts, Chicago, University of Illinois at Chicago, Chicago, IL\\
\texttt{agutfraind.research@gmail.com}}
\affil[2]{Institute of Theoretical Informatics,
Karlsruhe Institute of Technology (KIT),
Karlsruhe, Germany\\
\texttt{\{michael.hamann,meyerhenke,christian.staudt\}@kit.edu}}
\affil[3]{School of Computing,
              Clemson University,
              Clemson, SC\\
              \texttt{isafro@clemson.edu}}

\date{}

\begin{document}

\maketitle

\begin{abstract}
Research on generative models is a central project in the emerging field of network science, and it studies how statistical patterns found in real networks could be generated by formal rules. 
Output from these generative models is then the basis for designing and evaluating  computational methods on networks, and for verification and simulation studies. 
During the last two decades, a variety of models has been proposed with an ultimate goal of achieving comprehensive realism for the generated networks.  
In this study, we (a) introduce a new generator, termed \recon; (b) explore how \recon and some existing models can be fitted to an original network to produce a structurally similar replica, (c) use \recon to produce networks much larger than
the original exemplar, and finally (d) discuss open problems and promising research directions. 
In a comparative experimental study, we find that \recon is often 
superior to many other state-of-the-art network generation methods. 
We argue that \recon is a scalable and effective tool for modeling a given network while preserving important properties at both micro- and macroscopic scales, and for
scaling the exemplar data by orders of magnitude in size. 

\end{abstract}

\section{Introduction}\label{context-and-contribution}

Networks are widely used to represent connections between entities, because they 
provide intuitive windows into the function, dynamics, and evolution of natural and man-made systems.
However, high-quality, large-scale network data is often unavailable because of economic, legal, technological, or other obstacles \cite{chakrabarti2006graph,brase2009modeling}. 
For example, the human contact networks along with infectious diseases spread are notoriously difficult to estimate, 
and thus our understanding of the dynamics and control of epidemics stems from models that make highly simplifying assumptions or simulate contact networks from incomplete or proxy data  
\cite{Eubank04,Keeling_Rohani_2008,Meyers05}. In another domain, the development of cybersecurity systems requires testing across diverse threat scenarios and validation across diverse network structures that are not yet known, in 
anticipation of the computer networks of the future \cite{dunlavy2009mathematical}. In both examples, the systems of interest cannot be represented by a single exemplar network, 
but must instead be modeled as collections of networks in which the variation among them may be just as important as their common features.
Such cases point to the importance of data-driven methods for synthesizing networks that capture both the essential features of a system and realistic variability in order to use them in such tasks as simulations, analysis, and decision making. 

A good network generator must meet two primary criteria: realism and diversity. The first, realism, needs to consider any properties of the network that govern the domain-specific processes of interest such as system function, dynamics, and evolution. Hence, realism may depend on both structural network features and the more subtle emerging features of the network.  For example:
\begin{itemize}
\item Models of social networks should be able not only to reproduce structural features such as small-world properties, but also, and perhaps more importantly, to emulate emergent sociological phenomena such as  interactions between individuals in a community, as driven by their psychological needs and daily routines.  That is, the generated network should show similar interactions by its artificial individuals, as determined by implicit psychological and social rules.
\item Models of connected solar energy collectors of different sizes and capacities should simulate realistic energy outputs influenced by the weather.
\item Models of metabolic interactions should ultimately reflect biochemical properties of a cell. 
\end{itemize}

Second, a synthetic network should reflect naturally occurring stochasticity in a system, without systematic bias that departs from reality. This feature is important for benchmarking and evaluating the robustness of network-based algorithms, anonymizing networks, and generating plausible hypothetical scenarios. In particular, when engineering algorithms, the ability to create good synthetic test
data sets is valuable to estimate effectiveness and scalability of the proposed methods. 

In addition, a network generator should be effective in tasks such as obfuscation (replacing restricted real data with similar synthetic data), compression (storing only a generator and its parameters instead of large graphs), as well as extrapolation and sampling (generating data at larger or smaller scales).
Finally, the running time and memory requirements of the generator should be acceptable for realistically large datasets - datasets that may include millions or even billions of nodes and edges.

\subsection{Problem definition}
We envision the following usage scenarios. Given an original (or real) network $O = (V,E)$ ($n_o = |V|$, and $m_o = |E|$) that cannot
be freely shared, we would like to be able to create a synthetic network
$R$ (with $n_r$ nodes) that matches the original in essential structural
properties, so that computational results obtained from processing this
network are representative for what the original network would yield. We
refer to $R$ as a \emph{replica}. We assume that whoever creates the
replica has access to $O$ and can pass it to a \emph{model
fitting} algorithm which uses it to parametrize a generative model.

More importantly, in addition to producing \emph{scale-1 replicas} (where
$n_r = n_o$), in the second scenario we want to use the generative model for
\emph{extrapolation}, i.e., we would like to parametrize it to produce  produce a
\emph{scaled replica} $R^x$ that has $n_r = x \cdot n_o$ nodes, where $x$
is called the \emph{scaling factor}. The structural properties of $R^x$
should be such that they resemble a later growth stage of the original
(also see Section~\ref{sec:scaling-real}).
This should enable users of the replica to extrapolate the behavior of
their methods when the network data is significantly scaled.

Finally, with respect to performance, we would like the generator algorithm and
implementation as well as the fitting scheme to be efficient enough to produce large data sets (on the
order of several millions of nodes and edges) quickly in practice.

\subsection{Related work}

A number of network generation methods have been developed, and these fall into two classes: generative models and editing methods (see surveys in \cite{goldenberg2010survey,chakrabarti2006graph,brase2009modeling,dunlavy2009mathematical}).
The first set of methods produces (by using such elementary operations as randomization and replication) networks from small initial seed networks (sometimes empty). The goal of such generation is to produce the  structure that matches real data in \emph{prespecified} properties, such as
the degree distribution \cite{Albert2002,Newman03thestructure,mahadevan2006systematic}, clustering~\cite{Bansal09}, and the number of small subgraphs.
These methods are attractive because they often produce networks with the desired features and are grounded in well-developed theory (e.g., \cite{erdos1960erg}).
Some of these graph generators mechanistically model network growth \cite{barabasi,krapivsky,Leskovec08}, whereas others incorporate domain-specific information such as geographic location \cite{Watts1998} and cyber networks topological properties~\cite{medina,dunlavy2009mathematical}. 

One of the most successful generative strategies is based on Kronecker graphs \cite{kronmodel} (including stochastic Kronecker graphs; see also related work \cite{chakrabarti2006graph,palla2010multifractal}). Graphs generated by this model preserve properties such as degree distribution, diameter, and some eigenvalues. Generative methods often describe an evolutionary process that can potentially lead to the original network; however, the probability that it will lead to the structure that is approximately isomorphic to the original one at some coarse-grained resolutions (but not necessarily at finest) is usually negligible. This makes generative methods ill-suited for studies such as simulations when one may need to work with systems that are similar 
to the original.
Several generative models admit fast generators and are thus in our focus.
Among those models are RMAT~\cite{chakrabarti2004r}, BTER~\cite{kolda2013scalable},
and Hyperbolic Unit Disk Graphs (HUDG)~\cite{Krioukov2010}.
However, a previous fitting scheme by Leskovec \etal~\cite{leskovec2007scalable} for RMAT graphs
is quite time-consuming already for medium-sized networks \cite{staudt16algorithms}.

The other class of network generators, graph editing \cite{mihail2003markov}, 
starts with a given (real or empirical) network and randomly changes its components until the network becomes sufficiently different than the original network. 
These are designed to introduce variability while preserving key structural properties. 
A multiscale network generator \musk \cite{gutfraind2015multiscale} is an editing approach that is able to produce highly realistic synthetic networks by applying a local editing at different coarse-grained resolutions. The approach is comparable to \recon in the quality of generated networks; however, its implementation \cite{musketeer-impl} requires further work to be applicable on very large networks.

\subsection{Outline and contribution}
In this paper we develop and evaluate a fast generator that focuses on creating realistic 
\emph{scaled} replicas of complex networks.
We point out in Section~\ref{sec:design-goals} which criteria we consider important
for calling a (scaled) replica realistic. In particular we conceptualize realism
in two ways: (i) matching an original graph in a set of important structural properties, 
and (ii) matching the running time behavior of various graph algorithms. Typically, being an important applied criteria, the last is not considered by most existing network generators.

Our new generator \recon, short for \emph{\underline{Re}plication of \underline{Co}mplex \underline{N}etworks} and described in Section~\ref{sec:lfr-descr}, uses 
and extends ideas of LFR, a generator used for benchmarking community detection 
algorithms.
Using the original degrees and a discovered community structure, we are able to capture a much more detailed signature of the network than a parametrization of the LFR generator.
In Section~\ref{sec:fitting} we discuss the generative models that we use for comparison
(RMAT, HUDG, and BTER are among them)  and develop model fitting schemes for them.

Our comparative experimental study in Section~\ref{sec:experiments} indicates that \recon performs overall
quite well and usually better than other generators in terms of realism.
We can also conclude that the \recon implementation is fast, as it is capable of creating realistic 
scaled replicas on the scale of $10^8$ edges in minutes. The \recon code is publicly available
in the open-source network analysis package \nwk~\cite{StaudtSM16networkit}. In Section \ref{sec:openprob}, we discuss open problems and promising research directions in generating realistic replicas of complex networks. This paper is based on the conference paper \cite{staudt2016generating}.

\section{Realistic Replicas}
\label{sec:design-goals}
We consider a generative model realistic if there
is high structural similarity between the synthetic graphs produced and
relevant real-world networks. 
It is neither our goal nor generally desirable to obtain an exact correspondence between original and replica.
First, this would exclude the use case of obfuscation. Secondly, obtaining an isomorphic graph
is rarely required for generalizable experiments. 
Note that we consider a single ``realism score'' for each model inappropriately reductionist. 
Rather, we quantify diverse aspects of realism in our experimental evaluation and leave it to
the reader to decide about their relative importance.

For 1-scale replicas (with the same size as the original),
we measure the similarity in terms of a set of commonly used metrics:
Sparsity (number of edges vs number of nodes); degree distribution (more precisely its Gini coefficient); maximum degree
as a proxy for the connectedness of hub nodes; average local clustering coefficient to measure the local presence
of triangles; diameter to monitor the small-world effect; number of connected components and number of communities
as additional non-local features.
These metrics cover both local and global properties and are deemed important characteristics of networks~\cite{Newman2010networks}.

\begin{figure}[tbp]
  \begin{minipage}{\textwidth}
    \centering
    \includegraphics[height=3.5cm]{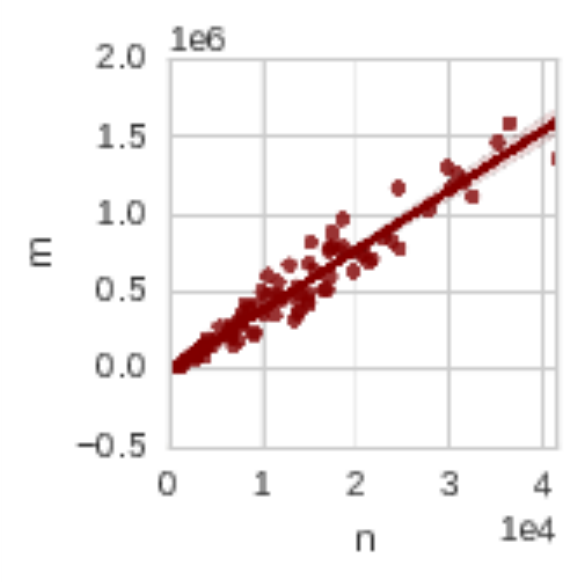}
    \includegraphics[height=3.5cm]{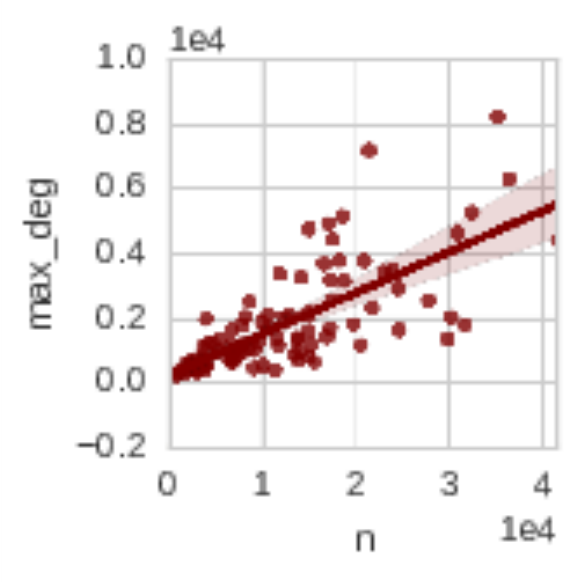}
    \includegraphics[height=3.5cm]{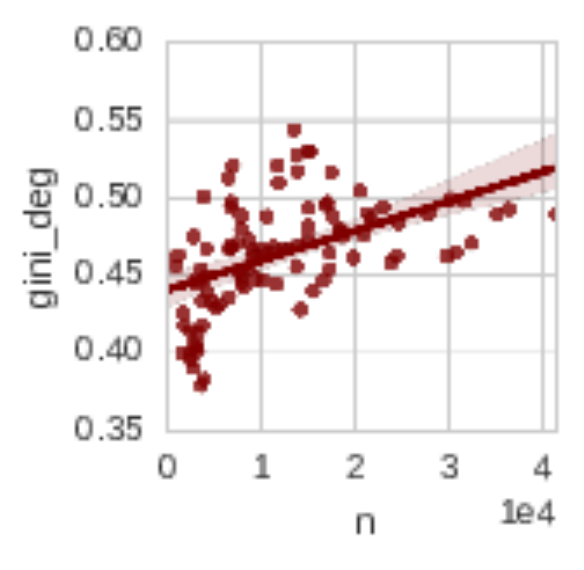}
    \includegraphics[height=3.5cm]{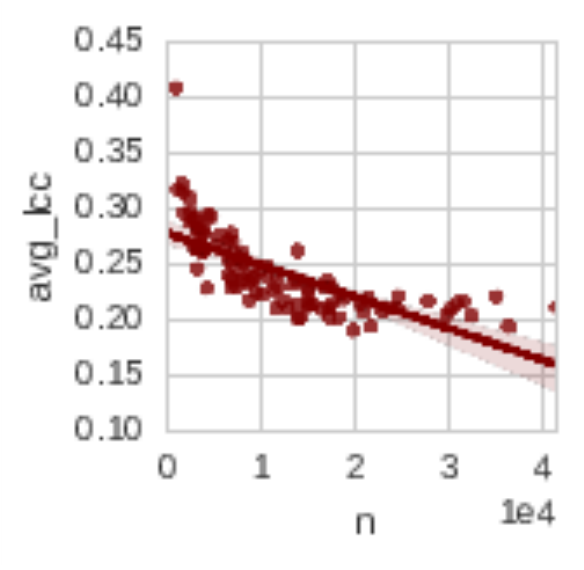}
    \includegraphics[height=3.5cm]{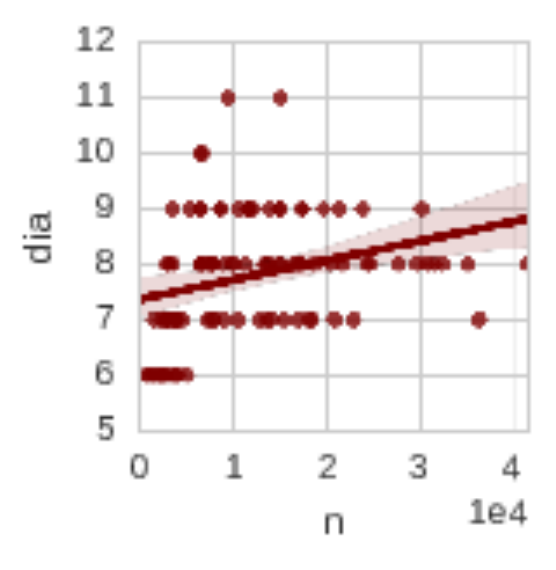}
    \includegraphics[height=3.5cm]{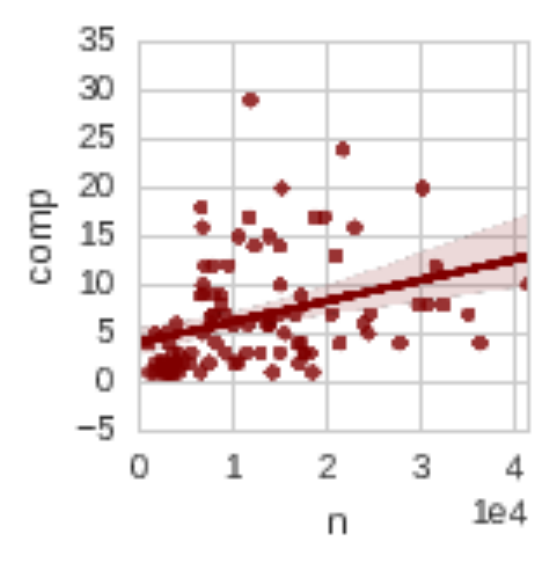}
    \includegraphics[height=3.5cm]{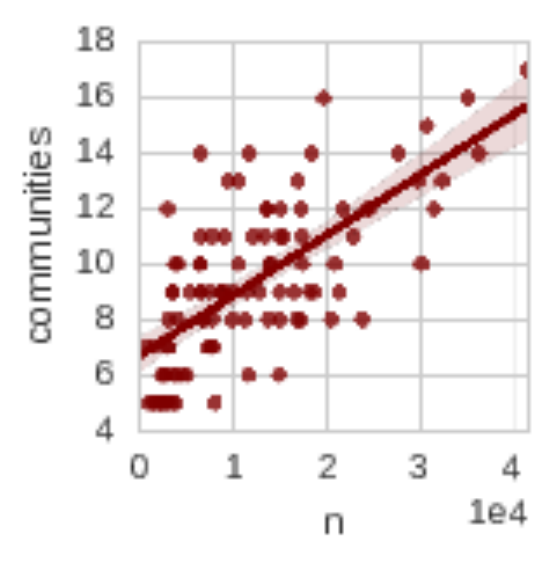}
  \end{minipage}
\caption{Scaling behavior of 100 Facebook networks; from left to right and top to bottom: number of edges, maximum degree, Gini coefficient of degree distribution, average local clustering coefficient, diameter, number of components, number of communities found by PLM}
\label{fig:fb-scaling}
\end{figure}

\label{sec:scaling-real}
How can we extend the notion above regarding realism to \emph{scaled} replicas 
of a network? To answer this question, let us look at the
scaling behavior of a set of 100 Facebook social networks~\cite{traud2012social}.
These networks were collected at an early stage of the Facebook online social networking service 
in which networks were still separated by universities.
In Figure~\ref{fig:fb-scaling} we demonstrate basic structural measures of these Facebook networks against
the number of nodes $n$, as well as a regression line and confidence intervals (shaded area) to emphasize
the trend. While linear regression may not always seem completely appropriate for these data, the general trend is typically still captured.

We can observe from Figure~\ref{fig:fb-scaling} a growth of the number of edges $m$ that is linear in $n$, an increase in the skew of the node degree
distribution as measured by the Gini coefficient, a growing maximum node degree, a slightly falling average local
clustering coefficient, a nearly constant
small diameter of the largest connected component, and a slightly growing number of connected components (which can be explained by some small connected components that exist in addition to a giant component).
We detect communities using \textsf{PLM} (Parallel Louvain Method), a modularity-based community detection 
heuristic~\cite{DBLP:journals/tpds/StaudtM16}, and report the number of communities minus the number of these small connected components. It can be observed that the number of non-trivial communities grows slightly.

While we do not propose that these scaling laws are universal, the trends represented here are commonly observed \cite{caldarelli2007large,boccaletti2006complex,snijders2001statistical}. Thus, we
use them to define desired scaling properties for the remainder of the
study as follows: $m$ grows linearly with $n$;  the diameter does not change significantly, preserving the ``small
  world property''; the shape of the degree distribution remains skewed; the maximum node degree increases; the number of connected components may grow; the number of communities increases slightly.%

Recall that one use case for our generator is testing of graph and network analysis algorithms. Since the 
running time is an essential feature in such tests, we also consider a realistic replication of running times
important. To this end, we select a set of graph algorithms that (i) compute important features of networks
and are thus frequently used in network analysis tasks and that (ii) cover a variety of patterns of computation and 
data access, each of which may interact differently with the graph structure. 
The set consists of algorithms for connected components (essentially breadth-first search),
PageRank (via power iteration), betweenness approximation (according to
Geisberger et al.~\cite{geisberger2008better}), community detection (\textsf{PLM}, \cite{DBLP:journals/tpds/StaudtM16}), core decomposition (according to~\cite{DBLP:conf/bigdataconf/DasariRZ14}), triangle counting (according
to~\cite{DBLP:journals/snam/HamannLMSW16}), and spanning forest
(essentially Kruskal's algorithm without edge weights).

\section{The Generation Algorithm \recon}
\label{sec:lfr-descr}

We introduce \recon{}, a generator for replicating and scaling complex networks.
Its input is a graph and a community structure on it.
For fitting a given graph without given community structure, we use \plm~\cite{DBLP:journals/tpds/StaudtM16} in order to detect a community structure first.
The basic idea of \recon{} is to randomize the edges inside communities and the edges between communities while keeping the node degrees.
This happens separately such that each community keeps as many edges as it had before.
For scaling a graph, we first create as many disjoint copies of the graph as desired and then apply the aforementioned steps.
During the randomization of the edges between the communities the copies usually become connected with each other.

The idea of randomizing graphs inside and between communities is inspired by the LFR generator, a benchmark graph generator for community detection algorithms~\cite{lancichinetti2008benchmark}.
There the basic building blocks are also a random subgraph per community and a global graph.
However, in the LFR generator the degrees and communities are not given but generated using a power law degree distribution and a power law community size distribution with nodes assigned to communities at random,  while \recon uses the given graph as input for them.

For randomizing graphs while preserving the degree sequence, we use random edge switches where two edges $\{u, v\}$, $\{y, z\}$ chosen uniformly at random are changed into $\{u, z\}$, $\{y, v\}$ if the resulting graph is still simple, \ie does not contain any duplicate edges or self-loops.
Similarly to the edge switching implementation provided by \cite{viger2005random} we use 10 times the number of edges as the number of random edge switches.
Previously performed experiments (\eg \cite{milo:2003}) have shown that this is enough to expect the resulting graph to be drawn uniformly at random from all graphs with the given degree sequence.

For an original graph $O = (V, E)$ with $n_o = |V|$ nodes and a desired scaling factor~$x$, \recon executes the following steps:

\begin{enumerate}
  \item Detect a community structure $\mathcal{C} = \{C_1, \dots, C_k\}$ on $O$ using \plm.
  \item Create $H$ as the disjoint union of $x$ copies of $O$. 
    The community structure is also copied such that the new community structure $\mathcal{D} = \{D_1, \dots, D_{x \cdot k}\}$ consists of $x \cdot k$ communities, \ie each copy of $O$ gets its own copy of the community structure that is aligned with the structure of the copied graph.
  \item For each community $D_i$, $1 \le i \le x \cdot k$, randomize the edges of the subgraph $H[D_i]$ that is induced by the community $D_i$ while keeping the degree distribution using random edge switches.
  \item Randomize the remaining edges, \ie all edges in $H$ that are not part of one of the subgraphs $H[D_i]$ using random edge switches.
    Note that afterwards some edges that were not in one of the $H[D_i]$ can now be inside a community.
    In order to avoid this, rewiring steps are performed by executing edge switches of such forbidden edges with random partners.
    A similar step is also used in the LFR generator, where it was observed that in practice only few rewiring steps are necessary \cite{lancichinetti2009benchmarks}.
\end{enumerate}

Note that it is not necessary to start with the full information of the original graph in Steps~3 and~4.
It is actually sufficient to know a community structure (as opposed to the whole original graph) and for each node the internal and external degree, \ie how many neighbors it has inside and outside its community, respectively.
For our implementation we choose this alternative.
Further, we execute Step~3 in parallel for all communities as the subgraphs are disjoint.

In addition to replicating important properties with high fidelity, the randomization in Steps~3 and~4 naturally produces random variance among the set of replicas.

\subsection{Fitting Generative Models to Input Graphs}
\label{sec:fitting}

Parametrized generative models require learning model input parameters
from the original network. A fitting scheme is an algorithm that takes a network as input and
estimates parameters of a generative model. Because, usually, such schemes are not unique, exploring them would be important future work. For this study, we have chosen one straightforward scheme per model, parameters of which are summarized in Table~\ref{tab:fitting-scaling}. Below we discuss a fitting scheme for power law degree distributions, and  briefly describe the generative models that are compared with \recon.\\

We consider an original graph $O = (V,E)$ with $n_o = |V|$, $m_o=|E|$ and
maximum degree $d_{\max}$.
We denote as $(a_i)_{i \in M}$ a sequence of
elements $a$ with indexes $i$ from an (ordered) set $M$, so
\[
\bigcup_{j = 0}^{k} (a^j_i)_{i \in M}
\]
denotes the ordered concatenation of $k$ sequences. The clustering coefficient of node $u$ is denoted by $\cc_u$. The number of nodes
with degree $d$ is denoted by 
\[
n_d = | V_d | = |\{v \in V: \deg(v) = d \} |,
\]
and 
\[
c_d = \frac{1}{n_d} \sum_{u \in V_d} \cc_u
\]
corresponds to the average clustering coefficient for nodes of degree $d$.

After a brief description of the considered models, we discuss fitting of power-law distributions, and continue with a discussion of the
parametrization of each model.

\begin{table*}[htb]
  \centering
\begin{small}
\begin{tabular}{p{1.6cm} p{1.8cm} p{4.8cm} p{4.8cm}}
{\bf Model}               & {\bf Parameters }                        & {\bf Fitting}  & {\bf Fitting}  Scaling by $x \in \mathbb{N}$ \\ \hline
Erd\H os--R\' enyi 	& $ER(n',p)$ 	& $n' = n$ \qquad $p = \frac{2m}{n \cdot (n-1)}$ 				& $n' = x \cdot n$ \qquad $p = \frac{2m}{x \cdot n \cdot (n-1)}$ 	\\ \hline
Barabasi-Albert	& $BA(n',k)$ 	& $n' = n$ \qquad $k = \lfloor m / n \rfloor$ 		& $n' = x \cdot n$ \qquad $k = \lfloor m / n \rfloor$	\\ \hline
Chung-Lu	& $CL(d)$ 	& $d = (\deg(u))_{u \in V}$ 		&  $d = \cup_{i=1}^{x} (\deg(u))_{u \in V}$	\\ \hline
Edge-Switching Markov Chain & $EMC(d)$ 	& $d = (\deg(u))_{u \in V}$ 		& $d = \cup_{i=1}^{x} (\deg(u))_{u \in V}$		\\ \hline
RMAT	& $RM(s, e, \newline (a,b,c,d))$ 	& $s = \lceil \log_{2} n \rceil$ \qquad $e = \lfloor m / n \rfloor$ \newline $(a,b,c,d) = \textsf{kronfit}(O)$		& 	$s = \lceil \log_{2} x \cdot n \rceil$  \qquad $e = \lfloor m / n \rfloor$ \newline $(a,b,c,d) = \textsf{kronfit}(O)$ \\ \hline
Hyperbolic \newline Unit-Disk & $HUD(n, \bar{d}, \gamma)$ 	& $n=n$ \qquad $\bar{d} = 2 \cdot (m / n)$ \newline $\gamma = \max\{2.1, \textsf{plfit}((\deg(u))_{u \in V})\}$		& $n=x \cdot n$ \qquad $\bar{d} = 2 \cdot (m / n)$ \newline $\gamma = \max\{2.1, \textsf{plfit}((\deg(u))_{u \in V})\}$	\\ \hline
BTER	& $BTER(d, c)$ 	& $d = (n_d)_{d \in (0, \dots, d_{\max})}$ \newline $c = (c_d)_{d \in (0, \dots, d_{\max})} $		&  $d = (n_d \cdot x)_{d \in (0, \dots, d_{\max})}$	\newline $c = (c_d)_{d \in (0, \dots, d_{\max})} $ \\ \hline
LFR	& $LFR(n',\newline \gamma,\bar{d},d_{\max}, \newline \beta, c_{\min},c_{\max})$	& 	$n' = n$ \newline $\gamma, d_{\min}, d_{\max} = \textsf{plfit}((\deg(u))_{u \in V})$ \newline $\zeta_s = \{ |C|\,|\,C \in \textsf{PLM}(O)\}$ \newline $\beta, c_{\min}, c_{\max} = \textsf{plfit*}(\zeta_s)$ & $n' = x\cdot n$ \newline $\gamma, d_{\min}, d_{\max}=\textsf{plfit}((\deg(u))_{u \in V})$ \newline $\zeta_s = \{ |C|\,|\,C \in \textsf{PLM}(O)\}$ \newline $\beta, c_{\min}, c_{\max} = \textsf{plfit*}(\zeta_s)$	\\ \hline

\end{tabular}
\end{small}
\caption{Parameters set to fit a model to a given graph, and to produce a scaled-up replica.}
\label{tab:fitting-scaling}
\end{table*}

\subsection{Considered Models}\label{sec:models}
Apart from \recon, we include the following models in our experimental study: \textit{Erd\H os--R\' enyi} random graphs (ER)~\cite{erdos1960erg}, %
\emph{Barabasi--Albert} (BA)~\cite{albert2002statistical}, %
\textit{Chung-Lu} (CL)~\cite{aiello2000random}, %
\emph{Edge-Switching Markov Chain} (ESMC), %
\textit{Recursive Matrix} (RMAT)~\cite{chakrabarti2004r}, %
\emph{Hyperbolic Unit-Disk Graph} (HUDG)~\cite{Krioukov2010}, %
BTER~\cite{Seshadhri2011},
and \emph{LFR}~\cite{lancichinetti2008benchmark}.%
\subsubsection{Fitting power law degree distribution (PLD)} 
Both LFR and HUDG generators produce graphs with a power
law degree distribution. Therefore, at least the power law exponent, and,
in the case of the LFR generator, also the minimum and maximum degrees
need to be determined such that the degree distribution fits the real
network. In Table~\ref{tab:fitting-scaling}, \textsf{plfit} refers to
our custom power law fitting scheme.

A practical replication of a network requires preserving the original average (otherwise, the density will be changed) as well as minimum and maximum degrees (applications can be sensitive to such fundamental properties as degree-1 nodes and the distribution
of hubs).

In general, it is assumed (and implemented in many algorithms~\cite{clauset2009power}) that PLD only holds from a certain degree on and that for smaller degrees, the distribution might be different.
As the LFR generator only generates a plain PLD, we cannot apply this assumption.
Therefore, we fit the PLD exponent such that, with the given minimum
and maximum degree, the average degree of the real network is expected
when a degree sequence is sampled from this PLD.
Using binary search in the range of $[-6, -1]$ we repeatedly calculate
the expected average degree until the power law exponent is accurate up
to an error of $10^{-3}$.\\

\subsubsection{Erd\H os--R\' enyi, Barabasi-Albert, Chung-Lu and
ESMC} 
\textit{Erd\H os--R\' enyi} random graphs (ER) \cite{erdos1960erg}, while not considered similar to most real-world graphs, are fundamental and an important baseline.%
The Erd\H os--R\' enyi model does not provide many options for
parametrization. The edge probability $p$ is set to produce the same
edge-to-node ratio $\frac{m}{n}$ as the original. The \emph{Barabasi--Albert} model (BA)~\cite{albert2002statistical} implements a preferential attachment process by which a PLD emerges, which has been claimed to be a typical feature of real complex networks. In BA,   we set the number of edges coming with each new node to fit the original edge-to-node ratio.
The \textit{Chung-Lu} (CL) model \cite{aiello2000random} recreates a given degree
sequence in expectation.
The \emph{Edge-Switching Markov Chain Generator} (ESMC) generates a graph that is randomly drawn from all graphs with exactly the given degree sequence (see
e.g.~\cite{milo:2003}, \cite{schlauch:2015}). In both CL and ESMC we use the original degree sequence. To generate larger networks, $x$ copies of this sequence are
concatenated, multiplying the number of nodes by $x$ while keeping the
relative frequency of each degree.\\

\subsubsection{RMAT} 
The \textit{Recursive Matrix} (RMAT) model \cite{chakrabarti2004r} was proposed
to recreate various properties of complex networks, including an
optional power-law degree distribution, the small-world property and
self-similarity. The RMAT model can only generate graphs with $2^s$ nodes, where $s$ is
an integer scaling parameter. In order to target a fixed number of nodes
$n_r$, we calculate $s$ so that $2^s > n_r$ and delete $2^s - n_r$
random nodes. The choice of the parameters $a,b,c,d$ requires some
discussion.

Leskovec et al. \cite{leskovec2007scalable} propose a method to ``given
a large, real graph {[}\ldots{}{]}, generate a synthetic graph that
matches its properties'', using stochastic Kronecker graphs: Starting 2-by-2 stochastic initiator matrix $I$,
Kronecker products are calculated so that $I^s$ is a stochastic matrix
of dimension $2^s$ that yields edge probabilities of a graph. This is
equivalent to the RMAT model as it yields the same edge probabilities.
They attempt to fit model parameters so that the likelihood that a given
original graph $O$ was generated starting from an initiator matrix $I$
is maximized, and propose the \textsf{kronfit} gradient descent
algorithm that iteratively estimates $I$ in $O(m)$ time. They do not
explicitly mention the case of creating a scaled replica, but it is
clear that the method is capable of producing graphs for arbitrary
exponents $s$. We use an implementation of \textsf{kronfit} which is
distributed with the \textsf{SNAP} network analysis tool suite~\cite{snap}.

\begin{figure}[ht]
\begin{center}
\includegraphics[width=0.6\textwidth]{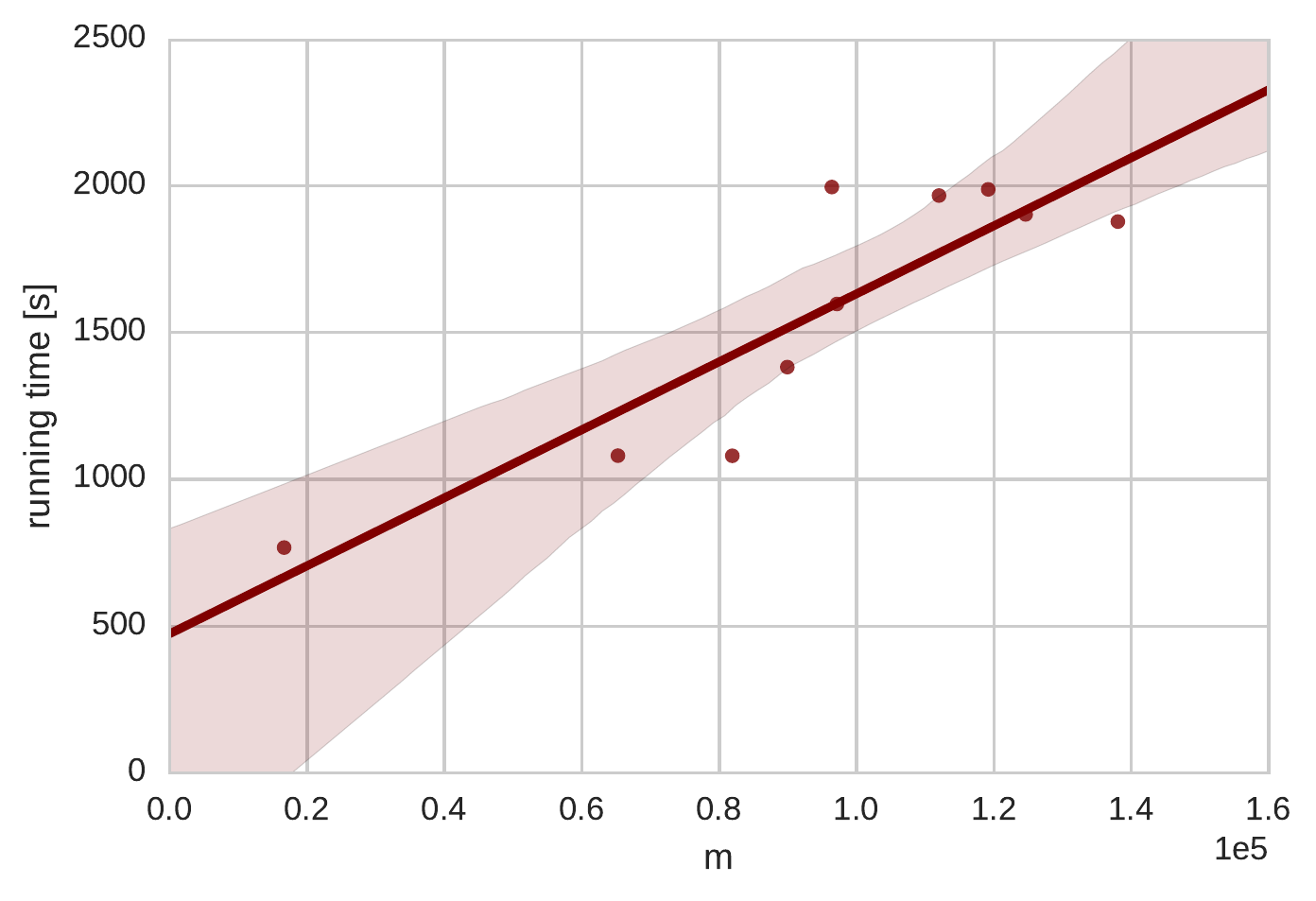}
\caption{Running time of 50 iterations of the \textsf{kronfit} algorithm in relation to the number of edges $m$ of the input network}
\label{fig:kronfit-scaling}
\end{center}
\end{figure}

The high running times of \textsf{kronfit} (see Figure \ref{fig:kronfit-scaling}) motivate a closer look
at the claim of increased realism as opposed to random guessing of the
initiator matrix. We compared properties of original networks, 
RMAT replicas with either random initator matrices or matrices generated with 50 iterations of
\textsf{kronfit} (a value proposed in
\cite{leskovec2007scalable}), see Appendix \ref{app:struc-prof}.
The most obvious difference between original and 
replicas is in the distribution of the local clustering coefficient, which is consistently lower for RMAT. 
Applying \textsf{kronfit} diminishes the mismatch slightly, but not fundamentally.
This points to inherent
restrictions of generating graphs by stochastic Kronecker
multiplication/RMAT that an elaborate parameter fitting does not
overcome.
Since running \textsf{kronfit} on every network to be replicated is not
practical, we estimate RMAT parameters as follows: We assume that the
100 Facebook networks constitute a class with essential structural
commonalities, and run the recommended 50 iterations of \textsf{kronfit}
on one typical network, \texttt{fb-Caltech36}. The resulting initiator is applied to replicate all Facebook networks. For other sets of
networks, the assumption of structural similarity cannot be made, so we
use a new random initiator for each replication.\\

\subsubsection{Hyperbolic Unit Disk Graphs (HUDG)} The random hyperbolic graph model embeds nodes into hyperbolic geometry and connects close nodes with higher probability~\cite{Krioukov2010}. The unit-disk variant HUDG we use in this paper connects
only nodes whose distance is below a certain threshold. We are focusing on the unit-disk variant to be
able to use a very fast generator for this model~\cite{DBLP:conf/isaac/LoozMP15}.
The model has been shown to replicate some properties observed in real networks, such as a power-law degree
distribution. This method receives as parameters the desired number of nodes, the average degree of the original network and a power law exponent which is fitted as described above. As the given power law exponent must be larger than 2, we supply at least an exponent of~$2.1$.\\

\subsubsection{BTER} The BTER model~\cite{Seshadhri2011} receives a degree distribution and the desired clustering coefficient per degree, i.e., for each degree to be realized the number of occurrences and the average clustering coefficient per degree.
For scaled replicas we scale the occurrences of all degrees by the scaling factor.
This leads to the target number of nodes while also preserving the general shape of the degree distribution.
In order to retain the distribution of the clustering coefficients, we leave them unchanged while scaling the network.\\

\subsubsection{LFR} The LFR was designed as a benchmark graph generator for community detection algorithms~\cite{lancichinetti2008benchmark}.
It generates graphs with a power law degree distribution and a community structure with community sizes that follow a power law distribution.
Apart from the number of nodes it requires parameters for power law distributions of the node degrees and the community sizes, and a mixing parameter that determines the ratio between intra- and inter-cluster edges.
We detect communities using \textsf{PLM} ~\cite{DBLP:journals/tpds/StaudtM16} and fit the parameters for the two power law distributions as described above using the original degree sequence and the found community sizes.
The mixing parameter is set to the ratio between intra- and inter-cluster edges of the found communities.
 As many networks in our set of real-world networks
contain a few small connected components, the smallest communities
usually only contain just 2 nodes while all other communities are much
larger. Therefore, frequently our minimum power law exponent 1 is chosen
as exponent and the expected average community size is still too low. In these
cases we also fit the minimum community size using binary search until
the expected average community size is the real average community size.
In Table~\ref{tab:fitting-scaling} this is denoted by \textsf{plfit*}.
For the scaling, we simply append copies of the same values.

\section{Computational Experiments}
\label{sec:experiments}
Our implementations of \recon{} and the various fitting methods are based on \textsf{NetworKit}~\cite{StaudtSM16networkit}, a tool suite for scalable network analysis. It also contains
many of the generators we use for comparison and provides a large set of graph algorithms we use for our experiments.
\textsf{NetworKit} combines C++ kernels with an interactive Python shell to achieve both high performance and interactivity, a concept we use for our implementations as well.
All implementations are freely available as part of the package at \url{https://networkit.iti.kit.edu}. 
This also includes a faster and parallel implementation of the LFR generator (compared to the original implementation \cite{lfrimpl}). 
\label{sub:exp-settings}
Our experimental platform is a shared-memory server with 256 GB RAM and 2x8 Intel(R) Xeon(R) E5-2680 cores at 2.7 GHz, using the GCC 4.8 compiler and the openSUSE 13.1 OS.
As described in Section~\ref{sec:design-goals}, we are interested in how well the different generators
replicate certain structural features of the original networks as well as the running times of various graph
algorithms. The results are described subsequently.
\subsection{Implementation details}\label{sec:implementation-appendix}
\label{lfr}
\begin{figure}[tb]
\begin{center}
    \includegraphics[height=5cm]{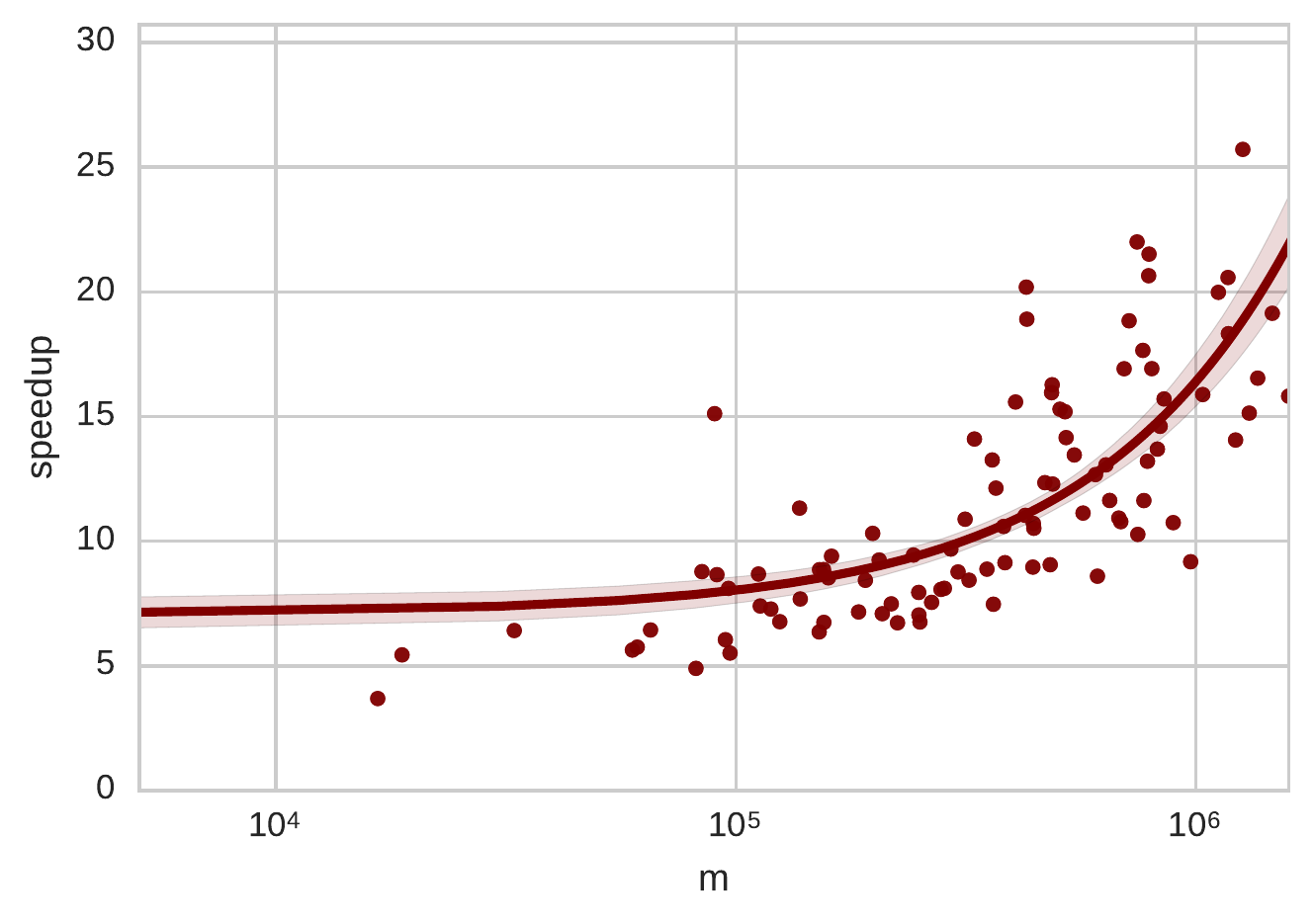}
\caption{Speedup of \nwk implementation of LFR compared to the reference implementation~\cite{lfrimpl} when replicating a set of networks with $m$ edges}
\label{fig:LFR-time}
\end{center}
\end{figure}

A reference implementation of the LFR generator by Fortunato \etal is
available online~\cite{lfrimpl}. We created a custom implementation of
LFR in \nwk, which is also the first parallelized implementation. 
Fig.~\ref{fig:LFR-time} illustrates the speedup that we achieve with a new implementation of the
LFR generator based on \nwk. Speed measurements were obtained on a machine with 16
physical cores. It yields speedup factors of 5 to 25 for the test set of
100 Facebook social networks, and the factor grows superlinearly with
the size of the network to be generated. 
The running time difference can be partially
traced back to implementation differences: The LFR reference
implementation relies on \texttt{std::set} to store and test for graph
adjacencies, while our implementation uses a simple but apparently more
efficient sequential scan on a simple array. Further, as mentioned
already, the community graphs are generated in parallel, which gives an
additional speedup.

\label{other-models}
For the tested generative models, \nwk includes efficient
implementations, such as implementations of linear-time algorithms
for the Erd\H os--R\' enyi and Barabasi-Albert models
\cite{batagelj2005efficient} and a subquadratic time algorithm for HUDG~\cite{DBLP:conf/isaac/LoozMP15}.
We have not implemented the BTER model; instead
we use the \textsf{FEASTPACK} implementation by Kolda et al.
\cite{feastpack}. \nwk implements an adapter class that performs the
model fitting and transparently calls the \textsf{MATLAB}-based
\textsf{FEASTPACK} binary. 

\subsection{Scaling behavior of
the generators}\label{scaling-behavior-of-generators}
The following experiments consider the scaling behavior of generative
models. Given the parametrization discussed before, we look at the
evolution of structural features with growing scale factor $x$ up to
$x = 32$. We consider the same basic scalar features as for the real networks in Section~\ref{sec:scaling-real}. %

\begin{figure}[tb]
\begin{center}
    \includegraphics[height=4cm]{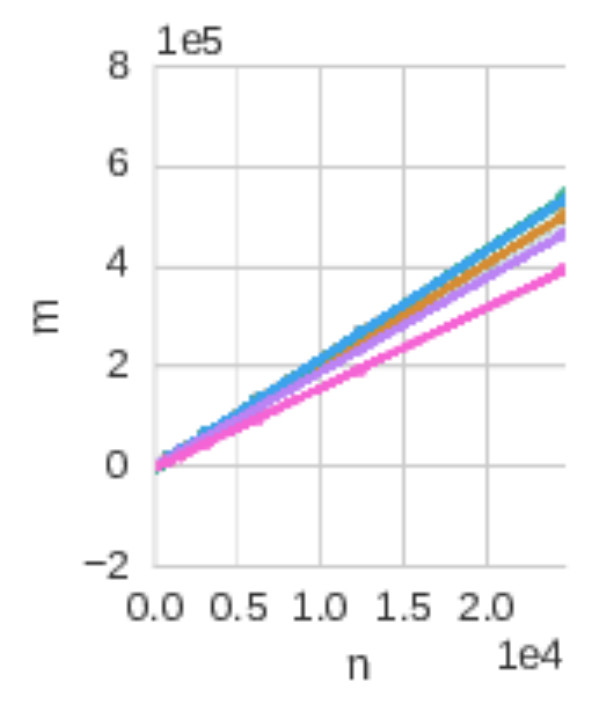}
    \includegraphics[height=4cm]{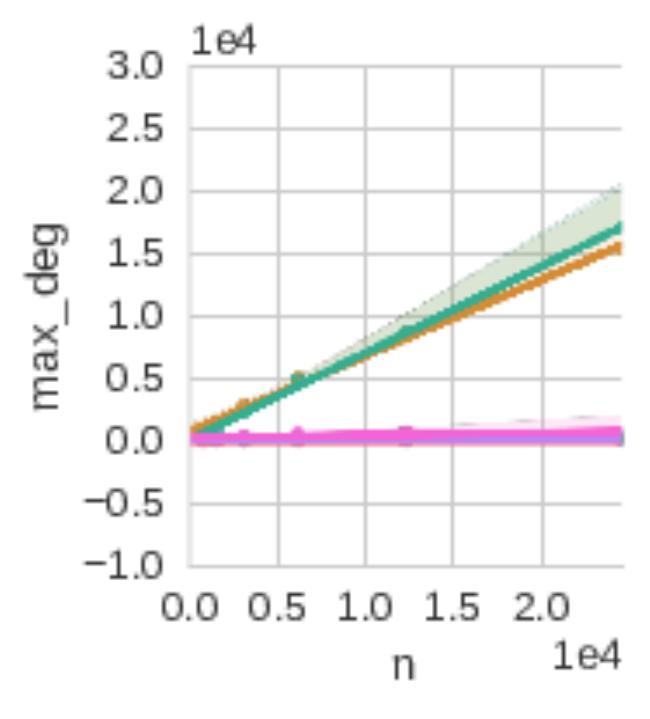}
    \includegraphics[height=4cm]{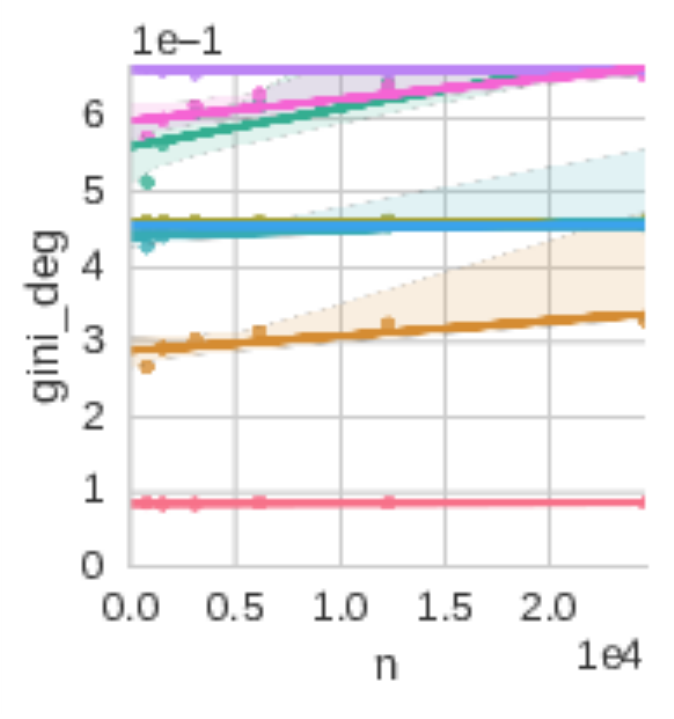}
    \includegraphics[height=4cm]{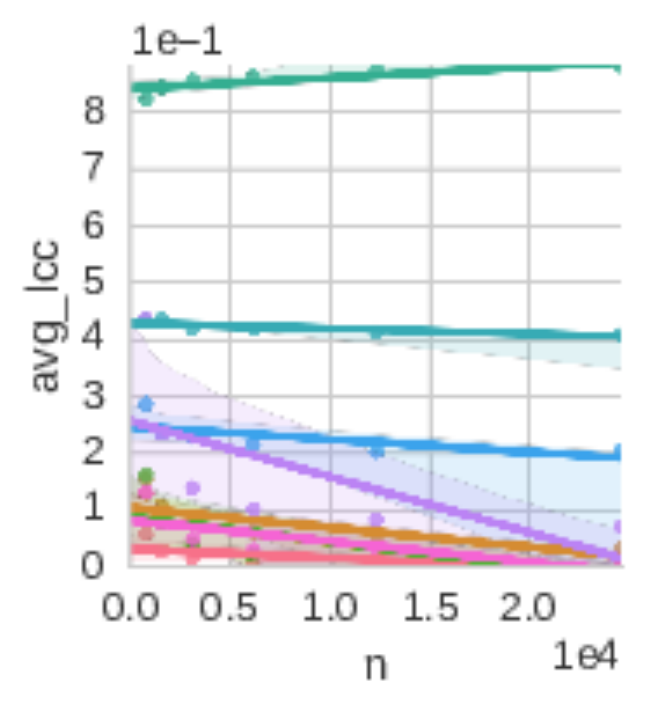}
    \includegraphics[height=4cm]{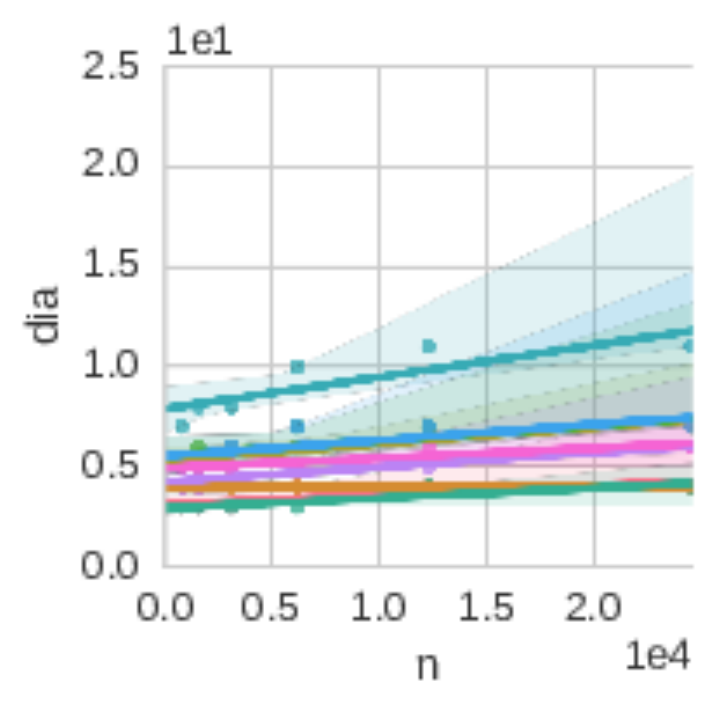}
    \includegraphics[height=4cm]{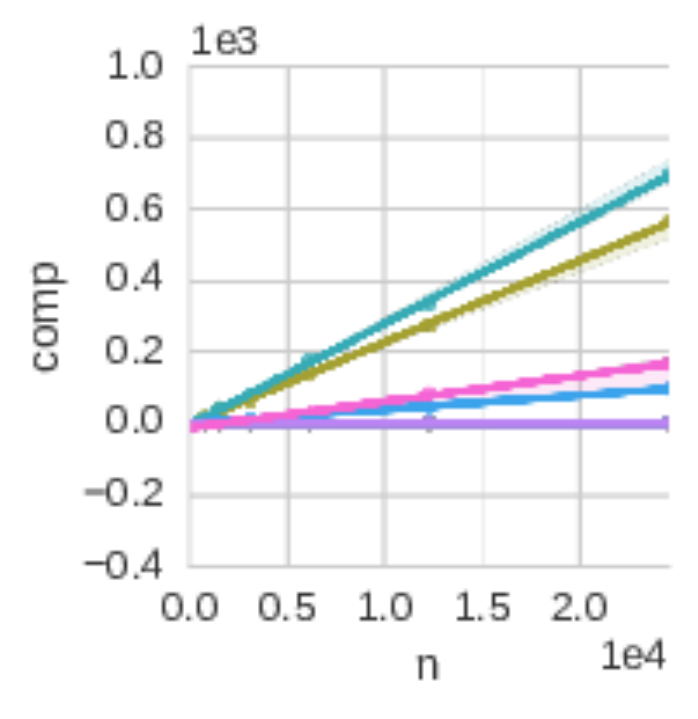}
    \includegraphics[height=4cm]{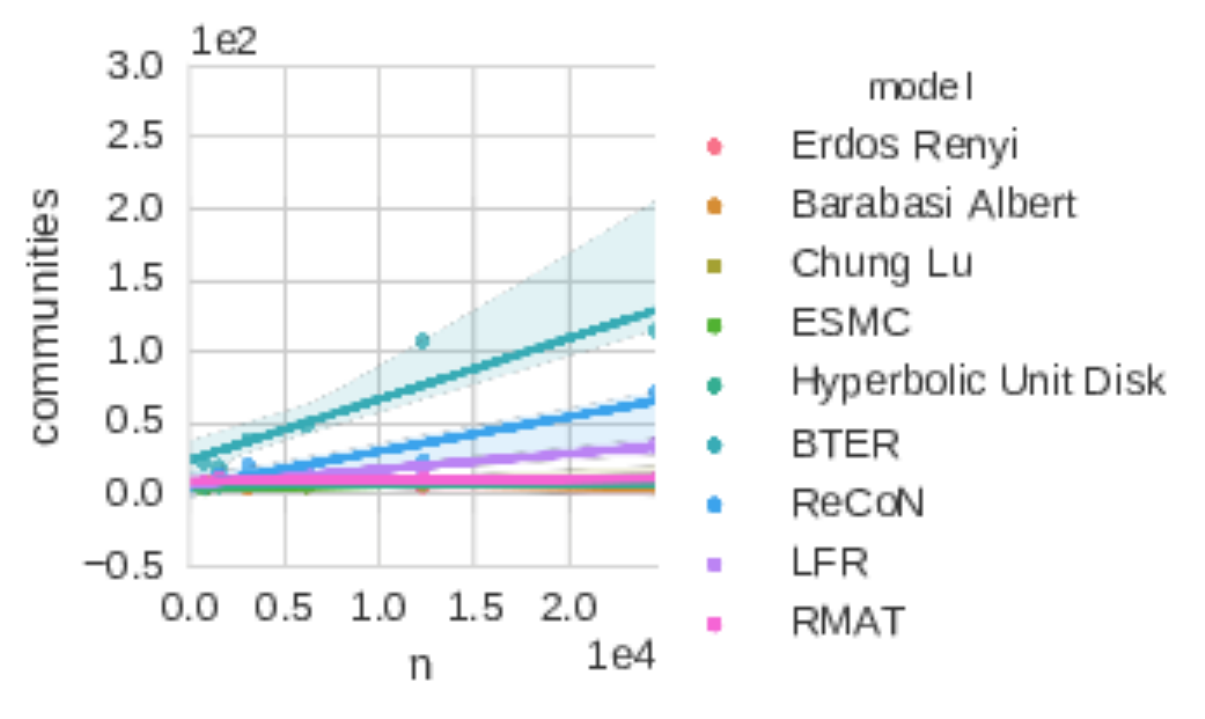}
\caption{Scaling behavior of the different generators on the \texttt{fb-Caltech36} network. From left to right and top to bottom: number of edges, max. degree, Gini coefficient of the degree distribution, average local clustering coefficient, diameter, number of components, number of communities.}
\label{fig:gen-scaling-fb}
\end{center}
\end{figure}

In Figure \ref{fig:gen-scaling-fb}, we show the results of the scaling
experiments for the \texttt{fb-Caltech36} network. The number of edges of
the replicas is increased almost
linearly by all generators to $\approx 5\cdot 10^5$ edges, which
approximately corresponds to 32 times the edges of the original network.
Therefore, all generators seem to keep the average degree of the original
network, which is expected as it is a parameter of all considered generators. %
Surprisingly, the maximum degree strongly increases up to 10 or 15 thousand with HUDG and BA generators, respectively. The
original maximum degree is 248, so that the new value is even significantly higher than the scaled
maximum degree (\ie 248 $\cdot$ 32).
Actually, from the scaling study in Sec.~\ref{sec:scaling-real}, we could expect an
increase, but rather in a lower range, so the degree distribution of 
BA and HUDG generators are not
realistic. Concerning the Gini coefficient, one can clearly see that ER
does not generate a skewed degree distribution at all. All generators
that get the exact degree sequence as input keep the Gini coefficient
constant, which is expected and also relatively realistic from our scaling study.

\begin{figure}[tb]
\begin{center}
    \includegraphics[width=\textwidth]{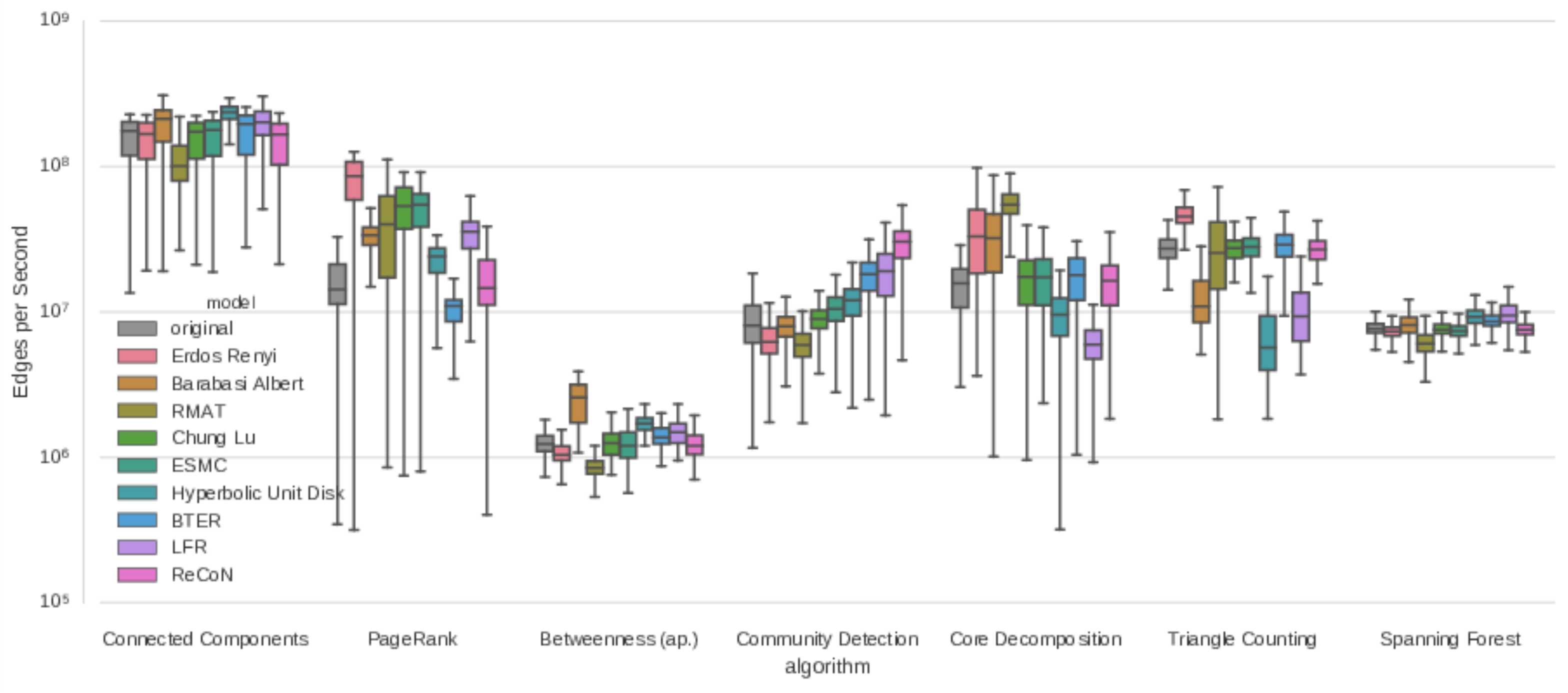}
    \label{fig:rt-replication-fb}
\vspace{-0.5cm}
\caption{Running time replication of a set of network analysis algorithms. Running times are in edges per second, i.e.,~higher is faster.}
\label{fig:rt-replication}
\end{center}
\vspace{-0.8cm}
\end{figure}

The original average local clustering coefficient of 0.43 is almost exactly reproduced by BTER in which it is an input parameter. The HUDG method increases it to 0.8, most others obtain very small values.
Our new \recon generator is less far off with 0.25 and a
slightly decreasing clustering coefficient; the latter is actually realistic
as we saw in Sec.~\ref{sec:scaling-real}. LFR is able to generate a clustering coefficient above 0.2 initially. 
Other generators produce much lower clustering coefficients. 
The original diameter of 6 is almost exactly kept by
\recon, all other generators except BTER produce networks with slightly
lower diameters, while BTER generates networks whose diameter is almost twice bigger. 
All generators show a slight increase of the diameter when the
networks are larger, which is consistent with our scaling study.
While most generators produce networks with just a single connected component, CL and BTER generate a large number, RMAT and \recon a moderate number of connected components.
In the case of CL, BTER and RMAT, this is probably due to a large number of degree-0 nodes.
The original network consists of a giant component and 3 small components; \recon scales them linearly, which is due to its parametrization.
The original network is split into eight non-trivial communities, that number should increase slowly according to Sec.~\ref{sec:scaling-real}.
Only in the networks generated by BTER, \recon and LFR, \textsf{PLM} can find a significant and increasing amount of communities.
While \textsf{PLM} finds over 100 non-trivial communities in the network generated by BTER, there are fewer communities detectable in the networks generated by \recon and even less in the ones generated by LFR.
Overall, \recon is the only generator that keeps the
degree distribution, and produces a realistic clustering coefficient and
a small diameter while keeping the graph connected and preserving a
moderate number of communities. All other generators are either unable
to keep the diameter or the connectivity or the number of communities.
It is part of future work to investigate whether the full hyperbolic random graph model 
can alleviate the weaknesses of the unit-disk case.

In Appendix~\ref{replicating-structural-properties} we present results of further experiments on the preservation of different network properties both for scale-1 and for scale-4 replicas.
The conclusion there is similar: for some properties, \recon is outperformed by other generators but overall it is able to produce the most similar replicas.

\subsection{Realism in the running times of graph algorithms}
\label{replicating-running-times}
Synthetic graphs are frequently used in algorithm engineering to
estimate the running time of an algorithm assuming that this time will be similar on real networks.
We examine if this is indeed the case with the generative models we consider.
Using the previously described generators and fitting schemes, we generate replicas of 100 Facebook networks and test a variety of graph algorithms (see Sec.~\ref{sec:design-goals}) on both
the original and replica sets. %

Our experiments demonstrate (see Fig. \ref{fig:rt-replication}) that the running times on the replica sets
often do not match those on the original set. 
The gray segments of the box plots represent
the distribution of running times measured on a set of original
networks. Ideally, the distribution on the synthetic networks would be
identical. The difference is statistically nontrivial, though.
Small variance between the models exists for connected components and
spanning forest computations, since their running time is nearly constant per edge.
Other algorithms exemplify how much running
time can depend on network structure, especially community detection,
core decomposition, triangle counting and PageRank. 
In general, the running time measurements obtained on \recon match the
originals closely in most cases. An exception is community
detection, where \plm seems to profit from \recon's explicit model of
communities. BTER shows close matches, too.

\subsection{Generator running times}
\begin{figure}
  \centering
\includegraphics[width=0.63\textwidth]{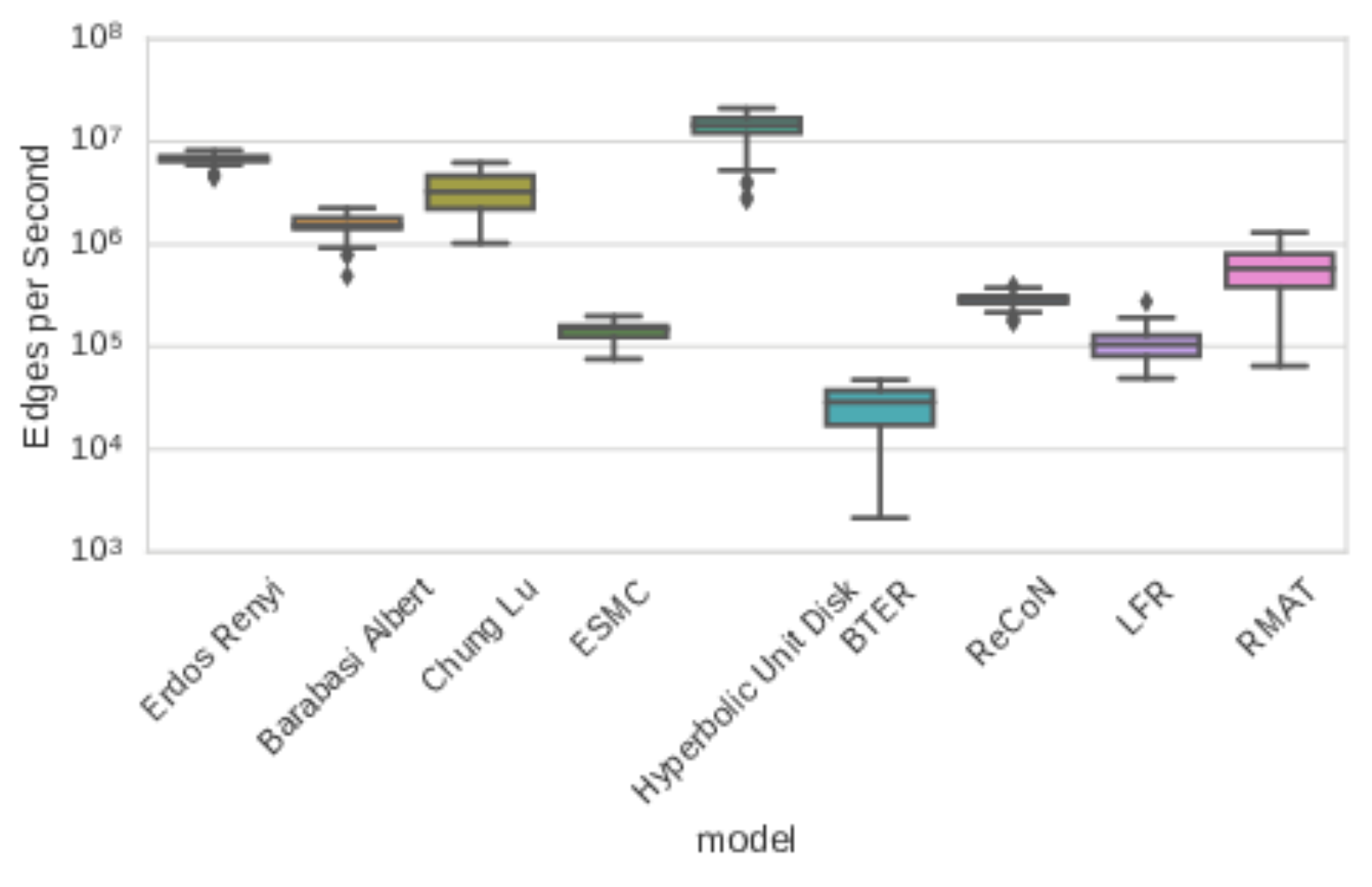}
\caption{Fitting and generating: processing speed measured in edges per second (size of replica graph/total running time, measured on 100 Facebook graphs)}
\label{fig:generation-speed}
\end{figure}
In Fig.\,\ref{fig:generation-speed}, we show the running times of parameter fitting and generating a replica for all methods. Processing
speed is given in the number of edges per second, a common measure for benchmarking graph algorithms that sets graph size and execution time in relation.
The entire set of Facebook networks was used to produce the measurements, so generated
replicas range from about 15000 to 1.5 million edges. For all models,
generating the graph takes up the vast majority of time. BTER's
\textsf{MATLAB}-based implementation is slowest, while the ER and HUDG
generators are the fastest. Our implementations of
LFR and \recon are not among the fastest generators, but fast enough to
produce millions of edges in minutes.

\subsection{Social Network: Example Replication I}\label{example-replication}

\begin{figure}[tb]
\begin{center}
\subfigure[original]{
	\includegraphics[width=0.22\textwidth]{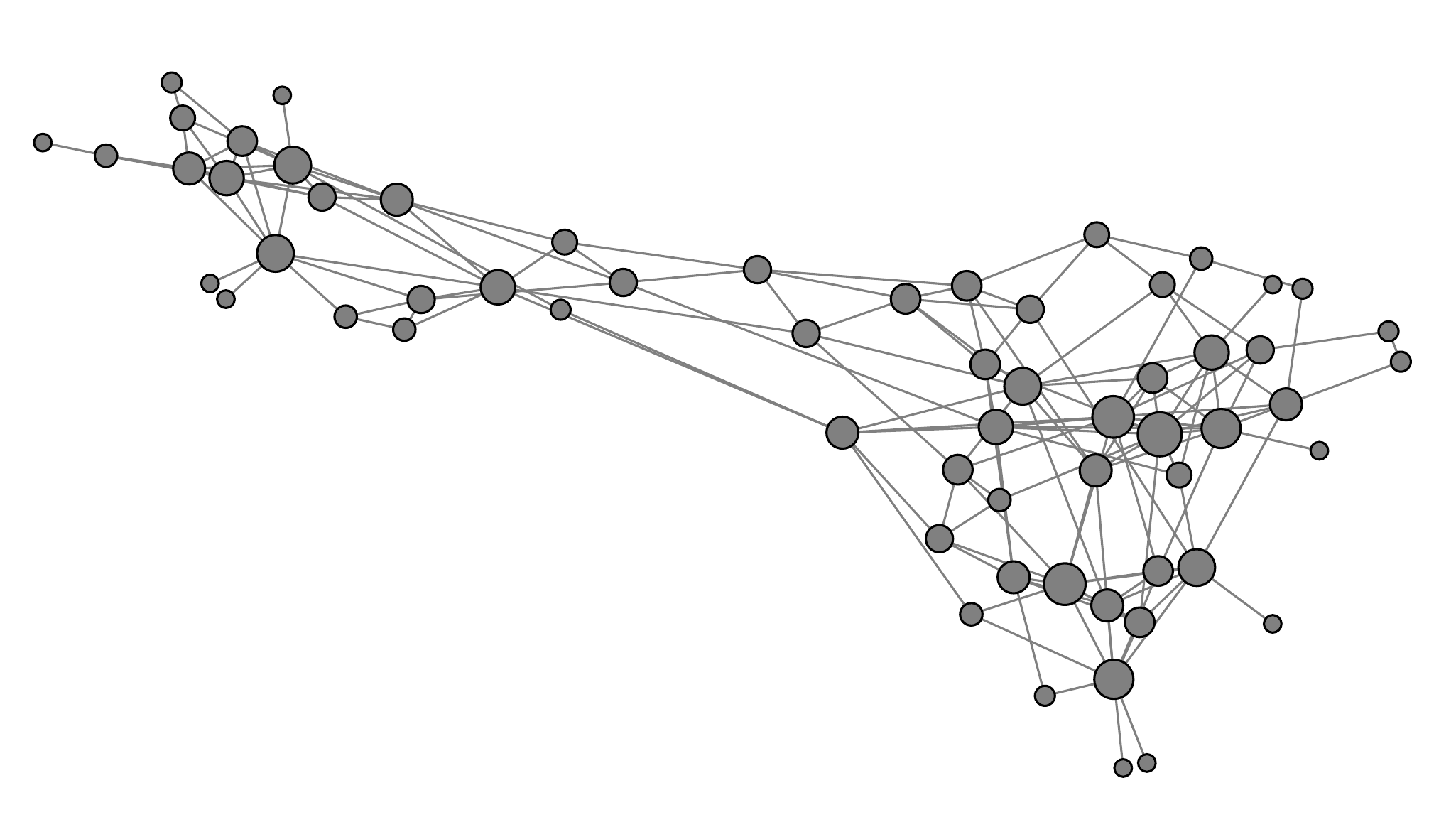}
	\label{fig:dolphins-original}
}
\subfigure[RMAT with \textsf{kronfit}]{
	\includegraphics[width=0.22\textwidth]{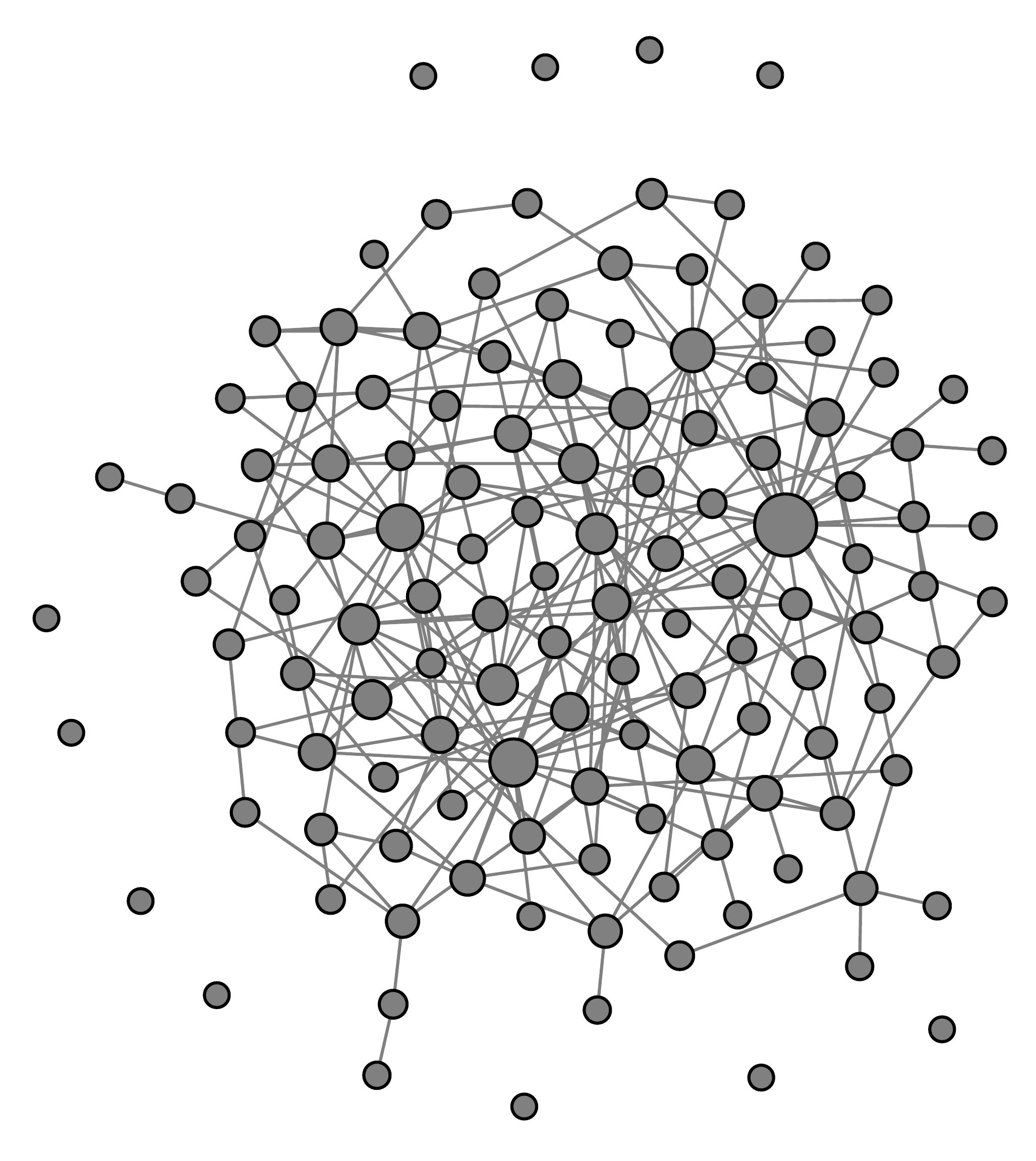}
	\label{fig:dolphins-rmat}
}
\subfigure[BTER]{
	\includegraphics[width=0.22\textwidth]{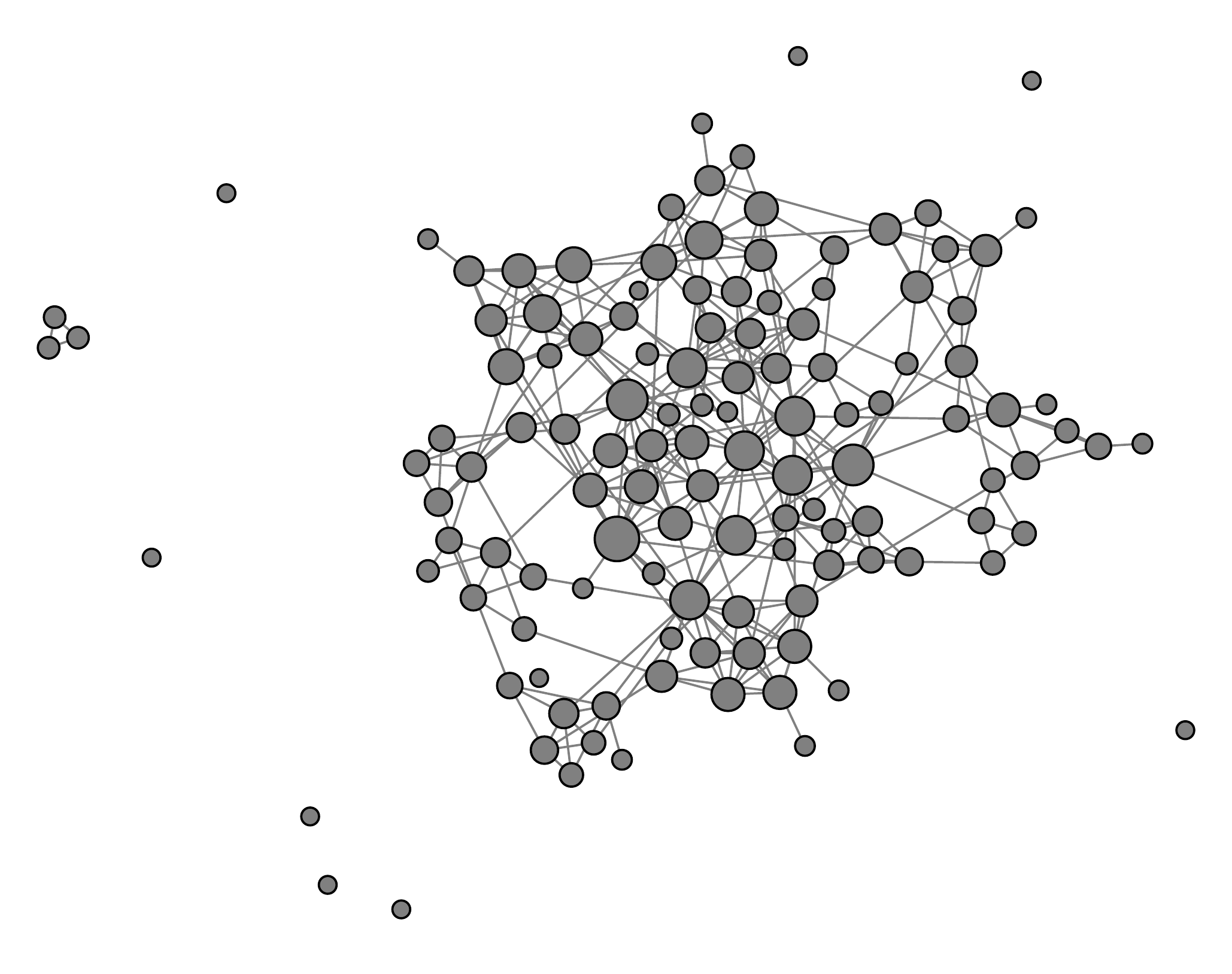}
	\label{fig:dolphins-bter}
}
\subfigure[\recon{}]{
	\includegraphics[width=0.22\textwidth]{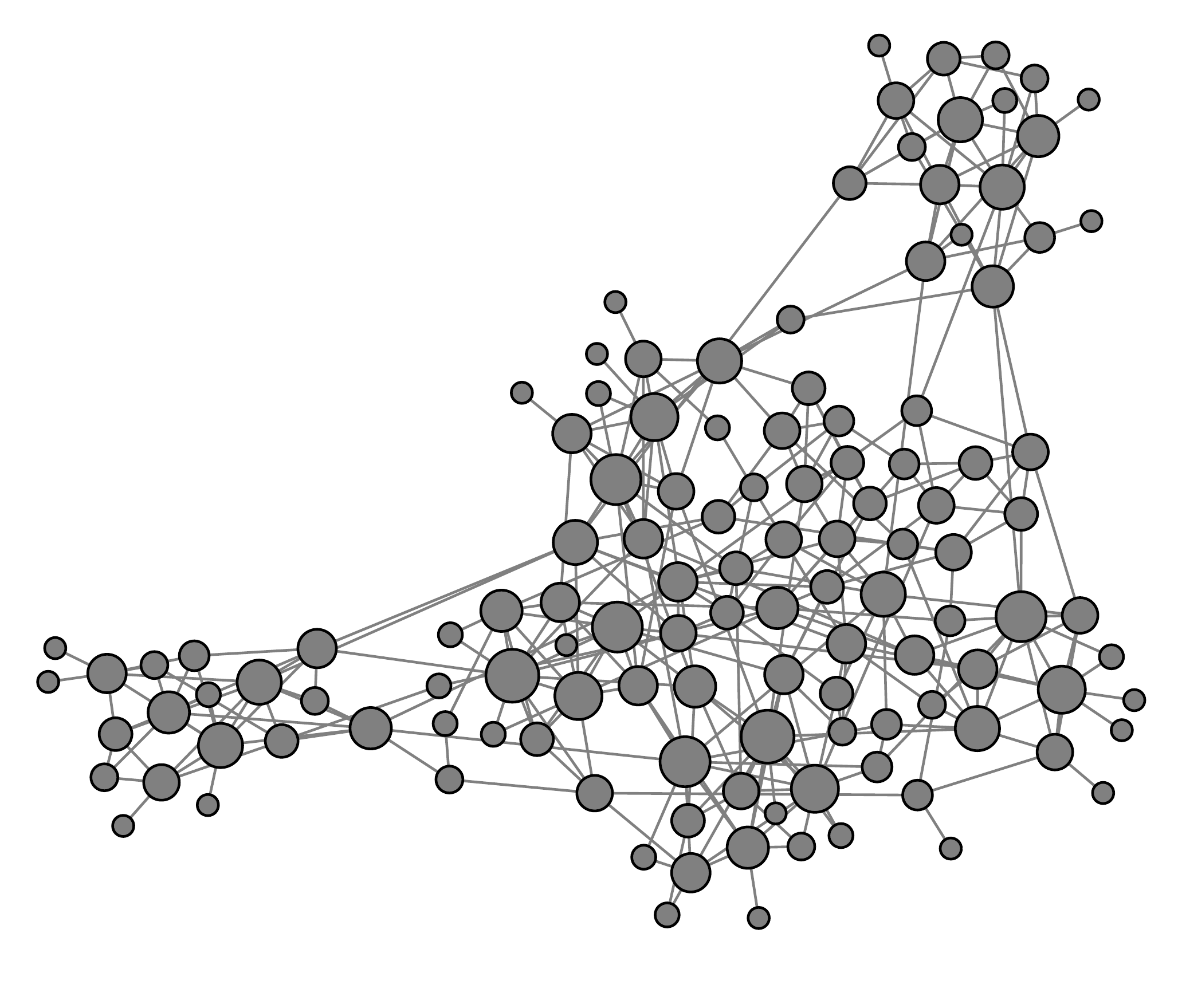}
	\label{fig:dolphins-lfr}
}
	\caption{A small social network and its scale-2 replicas produced by different models. \recon is the model that best reproduces a set of essential properties, including degree distributions, clustering and community structure.}
\label{fig:dolphins}
\end{center}
\end{figure}

As part of the evaluation of realism we replicate a small example
network with each of the models. The input graph is a social network of
Bottlenose dolphins. We use graph drawing with the \textsf{ForceAtlas2} layout
algorithm implemented in Gephi to inspect the network
structure. Node sizes are proportional to degree within
one drawing.
Fig.~\ref{fig:dolphins-original} shows a layout
of the original network.  Consider its basic structure: The graph is connected, with
two distinct communities separated by an edge bottleneck.
We also layouted replicas with scaling factors of 1 and 2 for each model, finding that the \recon replicas (e.g. Fig. \ref{fig:dolphins-lfr}) are the only
ones commonly recognized as resembling the original. BTER (Fig. \ref{fig:dolphins-bter}) is able to
generate a similar community structure, though not matching the original
as closely. All other replicas (e.g. those created by RMAT after applying \textsf{kronfit}, Fig. \ref{fig:dolphins-rmat}) lose the distinctive community structure
of the original.

\subsection{Epidemiological Network: Example Replication II}\label{sec:fitting-appendix}

\begin{figure}[tbp]
  \begin{center}
    \subfigure[original]{
      \includegraphics[width=5.5cm]{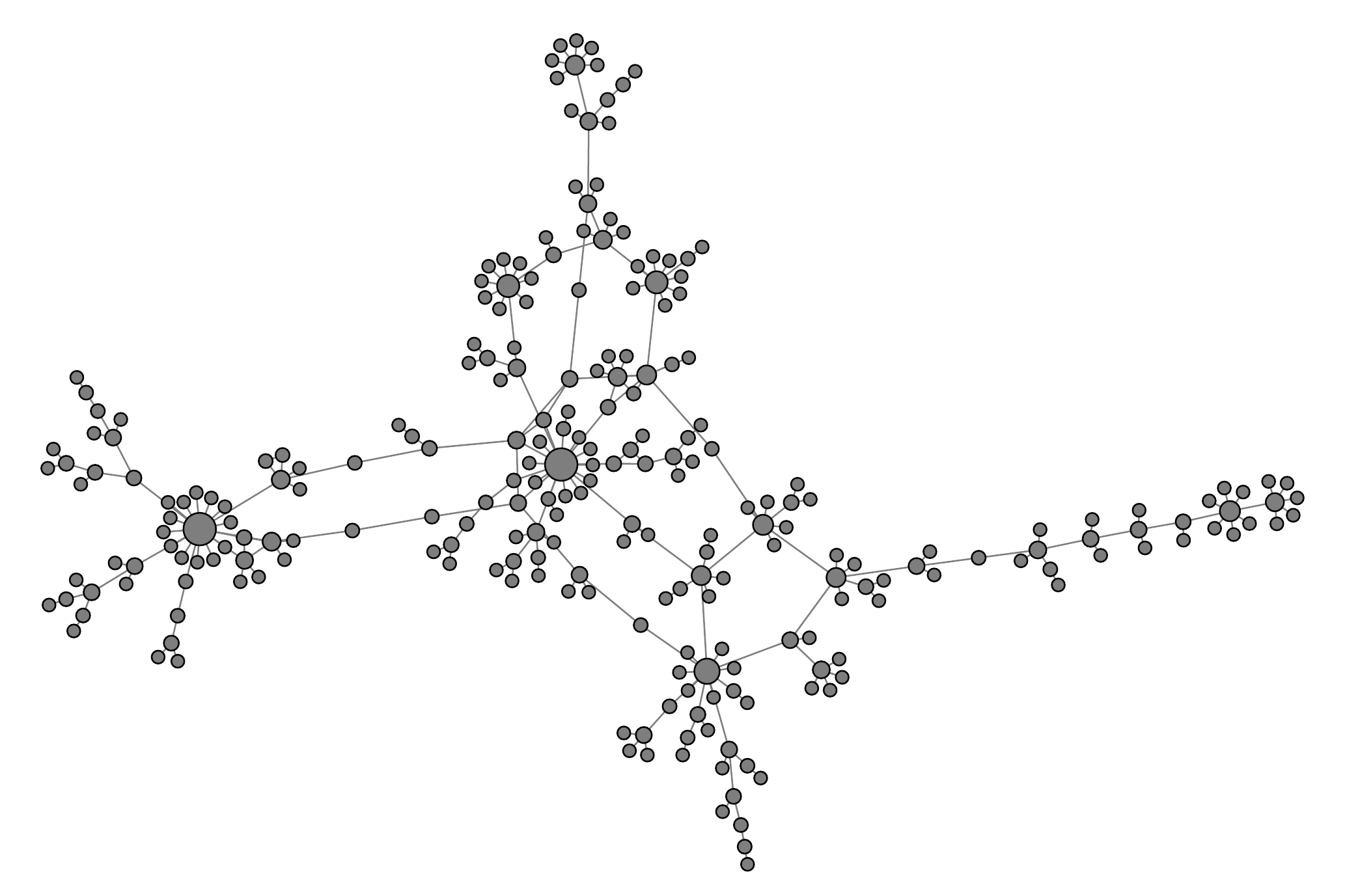}
      \label{fig:colorado-springs-orig}
    }
    \subfigure[\recon, scale-2]{
      \includegraphics[width=4.5cm]{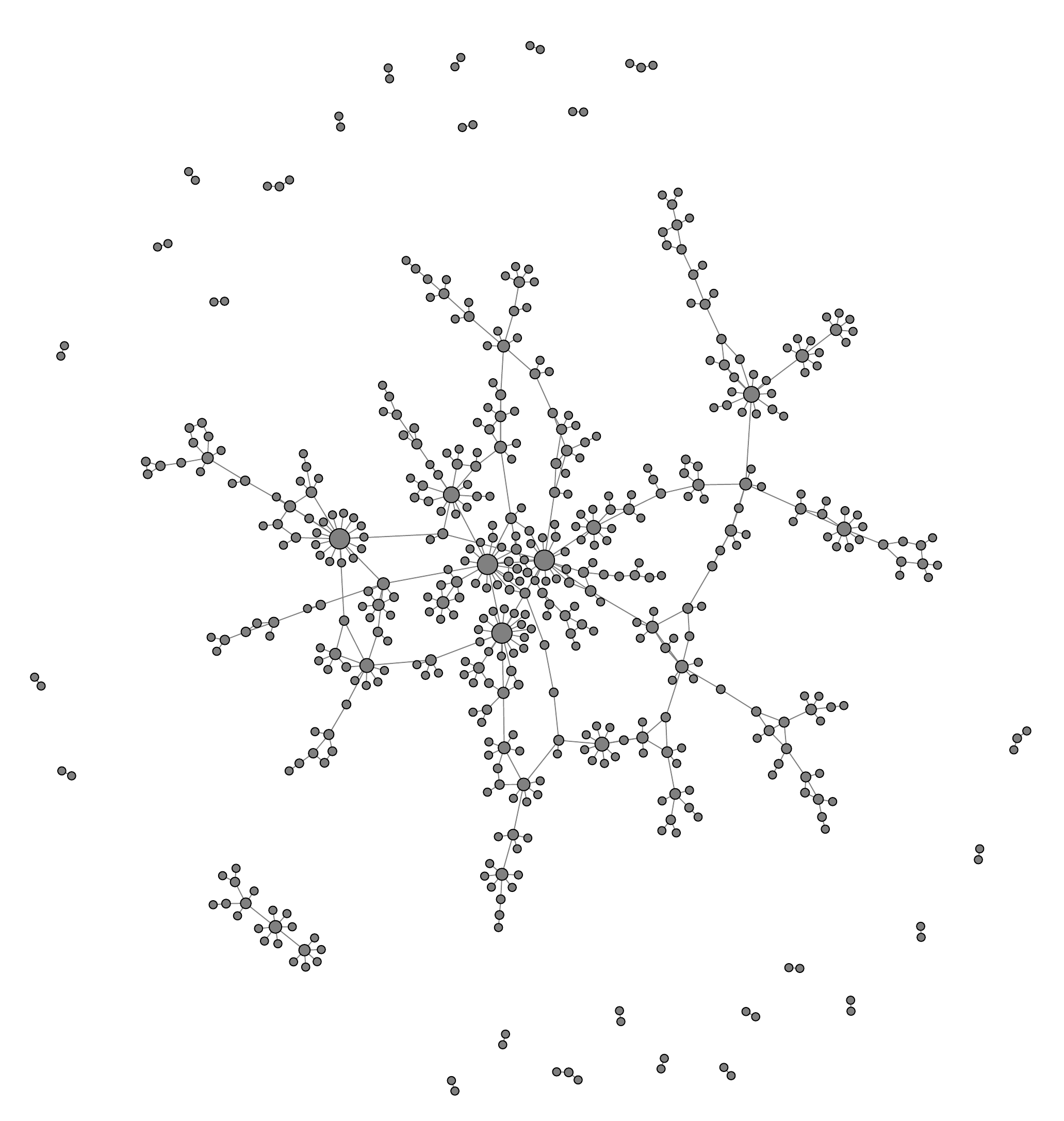}
      \label{fig:colorado-springs-scale-2}
    }
    \subfigure[\recon, scale-200k (sample)]{
      \includegraphics[width=3.5cm]{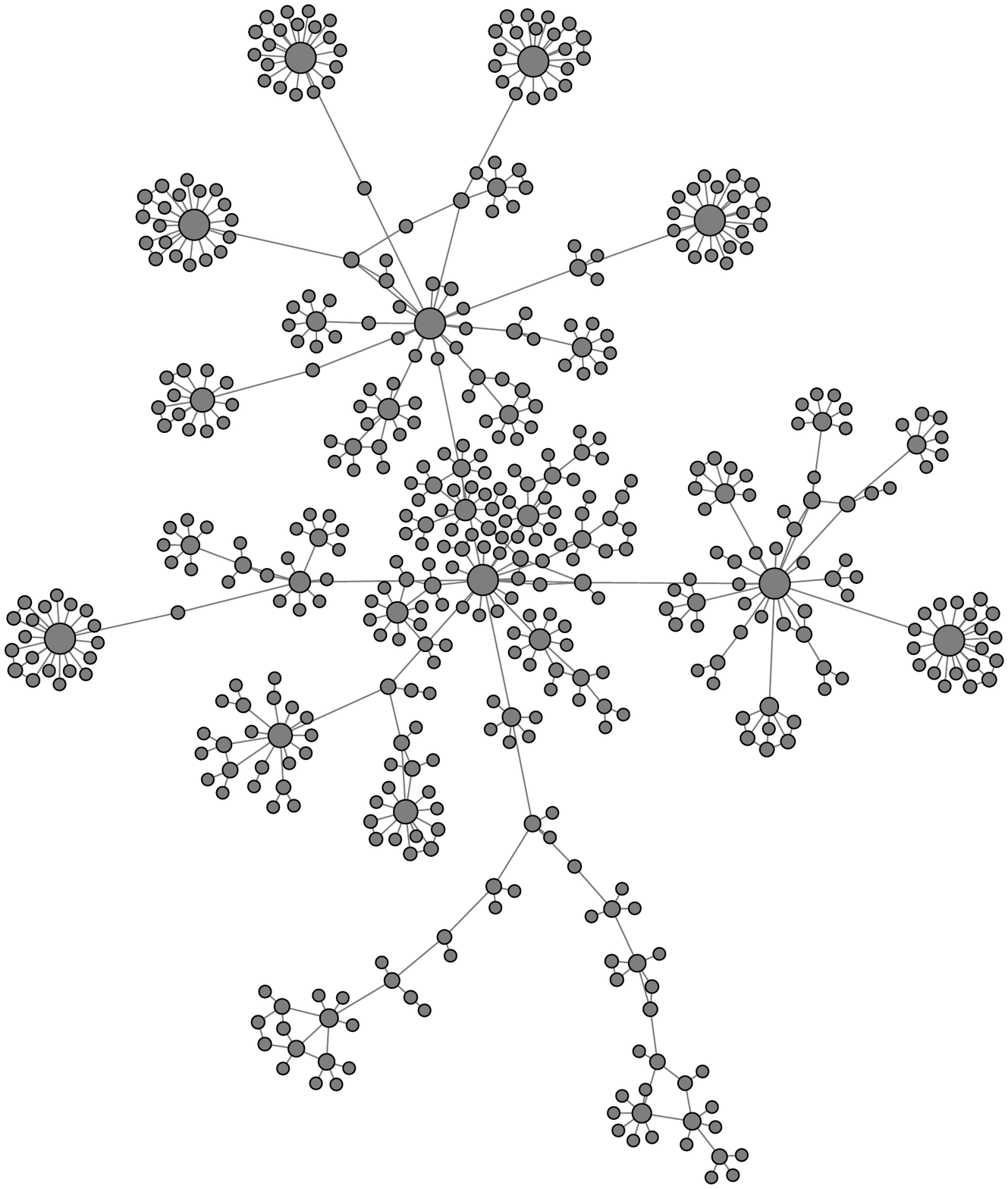}
      \label{fig:colorado-springs-scale-2k}
    }
  \end{center}
\caption{Colorado Springs epidemiological contact network}
\label{fig:colorado-springs}
\end{figure}

An epidemiological network frequently used in studies on the
transmission of HIV is based on data collected in Colorado Springs by
Potterat et al. \cite{potterat2002risk}. It contains 250 individuals
who were in contact in the 1980s through sex or injection drug use.
Figure \ref{fig:colorado-springs-orig} shows a force-directed layout
of the network's graph. Characteristic for this network is its tree-like
structure and the presence of high-degree hubs with attached
``satellite'' nodes of degree 1. The Colorado Springs network is an
instance of network data that cannot be shared freely due to legal
restrictions, but no such restrictions apply to a synthetic replica. A
replica made by the \recon generator reproduces structural features of the
original with the highest accuracy among the considered models. More
importantly, scaled replicas retain these essential properties,
including the hub-and-satellite structure. Figure~\ref{fig:colorado-springs-scale-2} shows the network replicated with
a scaling factor of 2, which is remarkably close to the original. Like
several other generators, \recon generates additional small connected
components as an artifact. If this is undesirable, a postprocessing step
could prune the network down to its giant connected component containing
a large majority of nodes. (Interestingly, the original data set
contained many singletons and isolated dyads that were removed from the
graph in a preprocessing step).

Real epidemiological contact network data is difficult to collect,
further complicated in the case of HIV by sex and drugs being tabu
subjects. This makes obtaining such a network on the scale of an entire
population impractical. In such a scenario, the ability to create
realistic large synthetic replicas of smaller real networks may be
highly relevant. As an explorative case study, we let \recon generate a
replica of the Colorado Springs network with 50 million nodes, which
corresponds to a scaling factor of 200 000. Figure~\ref{fig:colorado-springs-scale-2k} shows a sample from a
\(5 \cdot 10^7\) million node replica. This network's structure is quite
different from the clustered ``friendship networks'' of dolphins and
humans (Facebook). Remarkably, \recon also replicates many aspects of the
original's structure very closely, such as a tree-like structure with
hubs and attached satellites. These aspects are retained even for huge
scaling factors. This makes \recon a promising candidate to deliver large
data sets in cases where large amounts of real data are likely
unobtainable. Further domain-specific validation of the suitability of
such replicas would be interesting.

\section{Multi-scale modeling and open problems} 
\label{sec:openprob}
The \recon algorithm introduced above belongs to the class of single-scale 
generators in which the network structure is replicated in breadth in order to achieve an increased synthetic network without considering structural properties at coarse-grained resolutions of the network. 
That is, the community structure is replicated multiple times, and connections are randomized while preserving key properties of the original network. 
The approach allows us to extrapolate from a given network to the next level 
up; however, it is clear that as a network grows, its structure evolves new coarse-grained resolutions (or hierarchical levels) on top of those that were already present.

Progress in modeling such multiscale networks has been reported in  \cite{gutfraind2015multiscale}, which introduced a multiscale network generation approach, \musk. 
Briefly, starting from a single known or hypothesized network from any
domain, \musk synthesizes ensembles of networks that preserve, on average, a diverse set of structural features \emph{at multiple coarse-grained scales}, including degree, other measures of centrality, degree assortativity, path
lengths, clustering, and modularity, while introducing unbiased variability across the ensemble in many of these properties.
The core method is inspired by applications of the theory of multiscale methods   \cite{brandt:review01,briggs,mgbooktrott} to combinatorial optimization \cite{brandt:optstrat,DBLP:conf/gd/MeyerhenkeN015,safro:relaxml}. 
 The original computational multiscale method starts with an optimization problem represented on a network, 
 constructs a hierarchy of decreasing in size networks $G_0=O,G_1,...,G_k$ via a coarsening procedure \cite{safro:relaxml}, 
 solves a small optimization problem at the coarsest scale, and then iteratively uncoarsens the solution by gradual prolongations and local refinements from the coarse to next-finer scales.
 Similarly, in \musk a hierarchy of coarse networks is created; but, in contrast to the multiscale methods for computational optimization, nothing is optimized but the network is edited at all scales of coarseness. 
 During the editing process only local changes are allowed, which are the results of local decisions only. In other words, the problem of network editing/replication/randomization is formulated and solved at all scales, where primitives at the coarse scale (such as coarse nodes and edges) represent aggregates of primitives at previous finer scale. Analogous to multiscale methods for computational optimization problems  \cite{brandt:review01,brandt:optstrat}, by using appropriate coarsening, we are able to detect and use the ``geometry'' behind the original network at multiple scales, which can be interpreted as an additional property that is not captured by other network generation methods customized to replicate a small number of well known properties of $O$ such as degree distribution. 
 Moreover, it is known that the topology of many complex networks is hierarchical and thus might be produced through iterations of generative laws at multiple scales. In general, such generative laws often can be different at different scales, as evidenced by the finding that complex networks are self-dissimilar across scales \cite{Carlson02,Itzkovitz05,palla2005uncovering,Wolpert07,Binder08,Mones12}. These differences can naturally be reflected in the multiscale frameworks.

We believe that multiscale network modeling methods (MNM) are particularly promising for addressing
open problems in MNM, and we outline them below.
Some of the largest open problems are in the areas of statistical network hypothesis testing, network anonymization and compression, multiscale network decomposition and comparison, and understanding the fundamental limitations of network modeling.

\subsection{Statistical network hypothesis testing}
One of the biggest unsolved problems in network science is the problem of hypothesis testing.  
This is a problem of high significance in fields such as infectious diseases as well as other research and its application.
Given a network, the problem is to determine if the given network has statistically significant differences with a background network.
For example, given a security problem on a network, could we say that a given network is significantly more ``robust'' (or any macroscopic property) than what would be expected by chance, where ``chance'' is defined realistically?
The only existing approach, which is unsuitable for most applications, is to fit a parametric model such as an ERGM to the network, and then test for the parameters being different from $0$.
This ERGM approach cannot be used in the vast majority of applications, because the network cannot usually be represented realistically using an ERGM, nor could most properties be represented easily using a generative parameter.

However, given the high realism and versatility achieved by MNM methods, we believe that MNM, and non-parametric models more generally, could significantly advance statistical network hypothesis testing.
The non-parametric method promises to provide a much more realistic permutational test, and therefore identify anomalous structures in a given network at a much lower level of false positives.

\subsection{Network anonymization and compression}
A critical barrier towards advances in network science is the difficulty of sharing network data, which is often confidential and sensitive.
This is the problem of network anonymization and compression.
The multiscale strategy promises to address this gap through a powerful non-parametric data synthesis strategy, because, as we previously demonstrated, the strategy can generate networks that can model network structures with a high level of realism across a large spectral range. 
Initial success with such methods is possible already with the \recon and MUSKETEER strategies, but much work remains to be done addressing the spectrum of anonymization needs and developing versatile algorithms.
Current algorithms operate with complete network exemplars or replicas, but it could be beneficial for anonymization to operate with only small-size descriptions of the network such as motifs or possibly global parameters.

The MNM-based anonymization could be either non-parametric or parametric, allowing the user to limit the amount of data released to the public.  In the parametric version, the data owner shares a controlled amount of network information, while the recipient is able to obtain a realistic representation of the data by using an MNM algorithm.

Such a system would offer major advances in areas such as public health, cybersecurity and counter-terrorism.
Data in these areas is often very difficult to collect and/or locked by privacy laws.
Anonymization would compress the sensitive datasets and generate non-sensitive data, and in this form they could be shared with researchers in multiple institutions and even the public.
This would vastly expand the number of people who can perform research in critical areas of scientific and practical interest.

\subsection{Multiscale decomposition and comparison of networks arising from real-world complex systems}

The complexity of real network data suggests that its structure would be generally self-dissimilar when comparing several scales (or coarse-grained resolutions).
However, existing network generators generally fail to address the problem of self-dissimilarity. 
For example, the degree distribution or the clustering coefficient of a communication network are not necessarily the same as those at its backbone at macro- and meso-scopic scales. Moreover, in the vast majority of practical cases, the coarse-grained structure is not given (or can be found) explicitly, which makes the attempts to generate realistic synthetic network even more complicated. 
To address this problem, some of us and others have investigated a variety of multilevel optimization solvers for such problems, including graph partitioning \cite{bulucc2016recent,glantz2016tree,DBLP:journals/heuristics/MeyerhenkeS016}, 
linear ordering \cite{safro:relaxml}, graph drawing~\cite{DBLP:conf/gd/MeyerhenkeN015}, and node immunization \cite{leyffer2013fast} in which the multiscale organization of a graph is expressed via low-energy cuts. However, other advanced approaches are needed when a decomposition into scales is not cut-based.

A related open problem has to do with the comparison of networks with self-dissimilar properties.
In the existing multiscale algorithms like \recon and \musk, the comparison of the original and synthetic networks is done at the finest scale only. Whereas in some applications this can be acceptable, clearly, there are areas in which the multiscale structure is considered, and, thus, a multiscale comparison is much more beneficial and illuminating. However, to the best of our knowledge, there is no comprehensive concept of similarity that takes into account a comparison of properties at different scales.

\subsection{Fundamental limitation of network modeling algorithms}
Despite many studies of network modeling, we appear to lack a fundamental theory to understand the limitations of these algorithms.
There are at least three related problems.
First, we lack a theoretical understanding for many of the most popular generating algorithms, including multiscale strategies such as \musk.
Second, we lack a sense in which we can describe a network model or generator as in some sense ``optimal'' or ``most realistic''.
Perhaps it is possible to give an information-theoretic basis to understand the maximal realism a given generator can achieve given its parameter space.
Third, complex networks are typically associated with a variety of different processes running on or creating them. 
Nearly all network generation models focus on rules that are structuring the network, overlooking the close dependence between the structure-forming processes and the processes on the network. 
Indeed, in many domains, the realism of a network is closely tied to the dynamics on the network.
A domain expert (i.e., a potential user of network generation methods) cannot consider a synthetic graph realistic unless it becomes obvious that the processes that are supposed to co-exist with the network are also realistic (an issue we examined in the context of \recon). 
It is hoped that future studies in this field would examine this and earlier gaps in detail.

 \section{Conclusion}
 We have presented a new generator, \recon, for replicating and scaling
 existing networks. In an extensive experimental evaluation %
 we have shown that \recon is capable of generating networks which are (i) similar to the
 original network in terms of important structural measures and (ii) lead
 to similar running times of many graph and network analysis
 algorithms. Using \recon it is possible to realistically replicate an
 existing network, and to scale the synthetic version by orders of
 magnitude, \eg, in order to test algorithms on larger data sets where
 they are not available. Furthermore, it allows to create anonymized copies of such
 networks that can be distributed freely and allow to conduct representative
 experiments on them. While other generators sometimes perform better
 concerning certain criteria, none of the other generators is capable of
 approximately reproducing such a wide range of properties and running
 times.

\appendix

\clearpage

\section{Appendix: Structure profiles for the evaluation of the kronfit scheme}\label{app:struc-prof}

The high computational cost of \textsf{kronfit} motivates a closer look
at the claim of increased realism as opposed to random guessing of the
initiator matrix. Network profiles generated by \nwk enable a quick exploratory analysis of an
original network and replicas created by RMAT, with random parameters on
the one hand and parameters generated after 50 iterations of
\textsf{kronfit} (a value that the authors propose in
\cite{leskovec2007scalable}). The original is \texttt{fb-Caltech36}, a
small instance from the ``Facebook 100'' collection of social networks.
Figures \ref{fig:kronfit-profiles}, and \ref{fig:kronfit-profiles2} show an excerpt from the generated
profiles. The most obvious difference between the original and its two
replicas is in the distribution of the local clustering coefficient:
While the median is around 0.4 for the original, it is about 0.1 for the
replica with random parameters and only slightly higher for the
\textsf{konfit}-generated replica. The same behavior of the generator
was observed consistently on other networks. This points to inherent
restrictions of generating graphs by stochastic Kronecker
multiplication/RMAT that an elaborate parameter fitting does not
overcome.

Since running \textsf{kronfit} on every network to be replicated is not
practical, we estimate RMAT parameters as follows: We assume that the
100 Facebook networks constitute a class with essential structural
commonalities, and run the recommended 50 iterations of \textsf{kronfit}
on one typical network, \texttt{fb-Caltech36}. The resulting initiator

\[
I_{\mathtt{fb-Caltech36}} = 
\begin{bmatrix} 
0.378802757 & 0.249474498 \\ 
0.255098510 & 0.116624233 
\end{bmatrix}
\]

is applied to replicate all Facebook networks. For other sets of
networks, the assumption of structural similarity cannot be made, so we
use a new random initiator for each replication.

\begin{figure}[htbp]
  \centering
\subfigure[original: \texttt{fb-Caltech36}]{
    \includegraphics[width=0.8\textwidth]{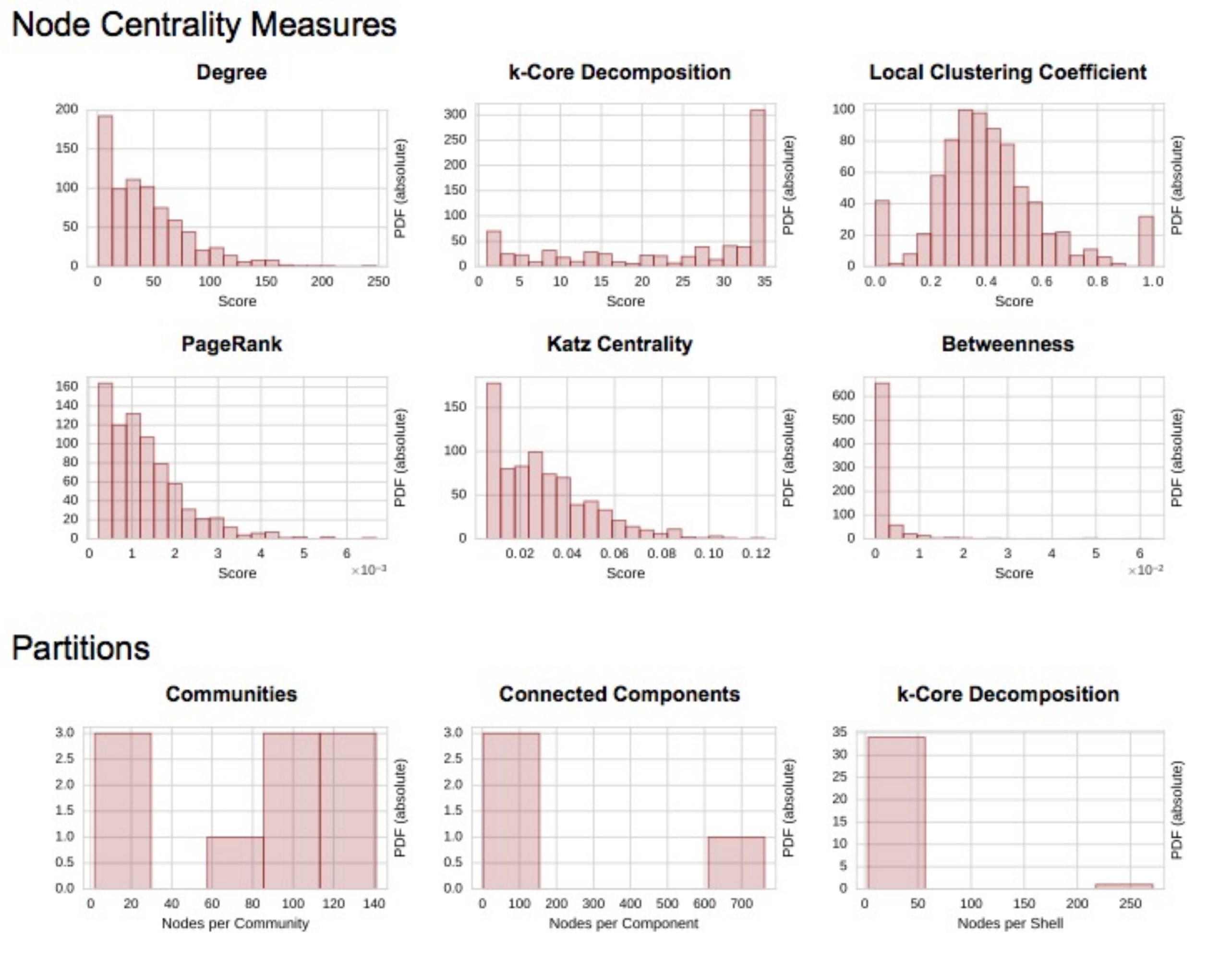}
}
\subfigure[RMAT replica with random initiator matrix]{
    \includegraphics[width=0.8\textwidth]{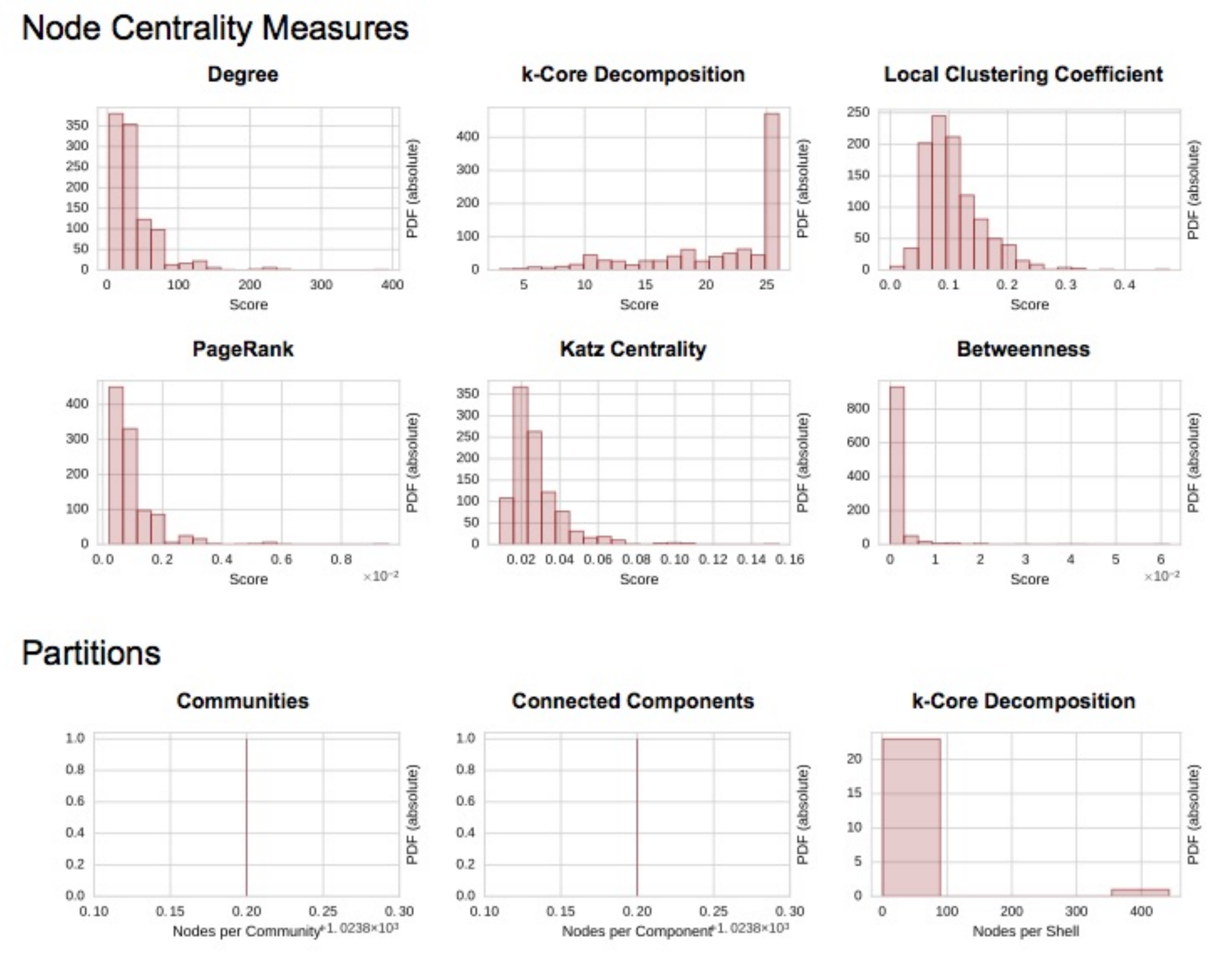}
}
\caption{Structure profiles for the evaluation of the \textsf{kronfit} scheme}
\label{fig:kronfit-profiles}
\end{figure}

\begin{figure}[htbp]
\begin{center}
\subfigure[RMAT replica with \textsf{kronfit}-generated initiator matrix]{
\includegraphics[width=0.8\textwidth]{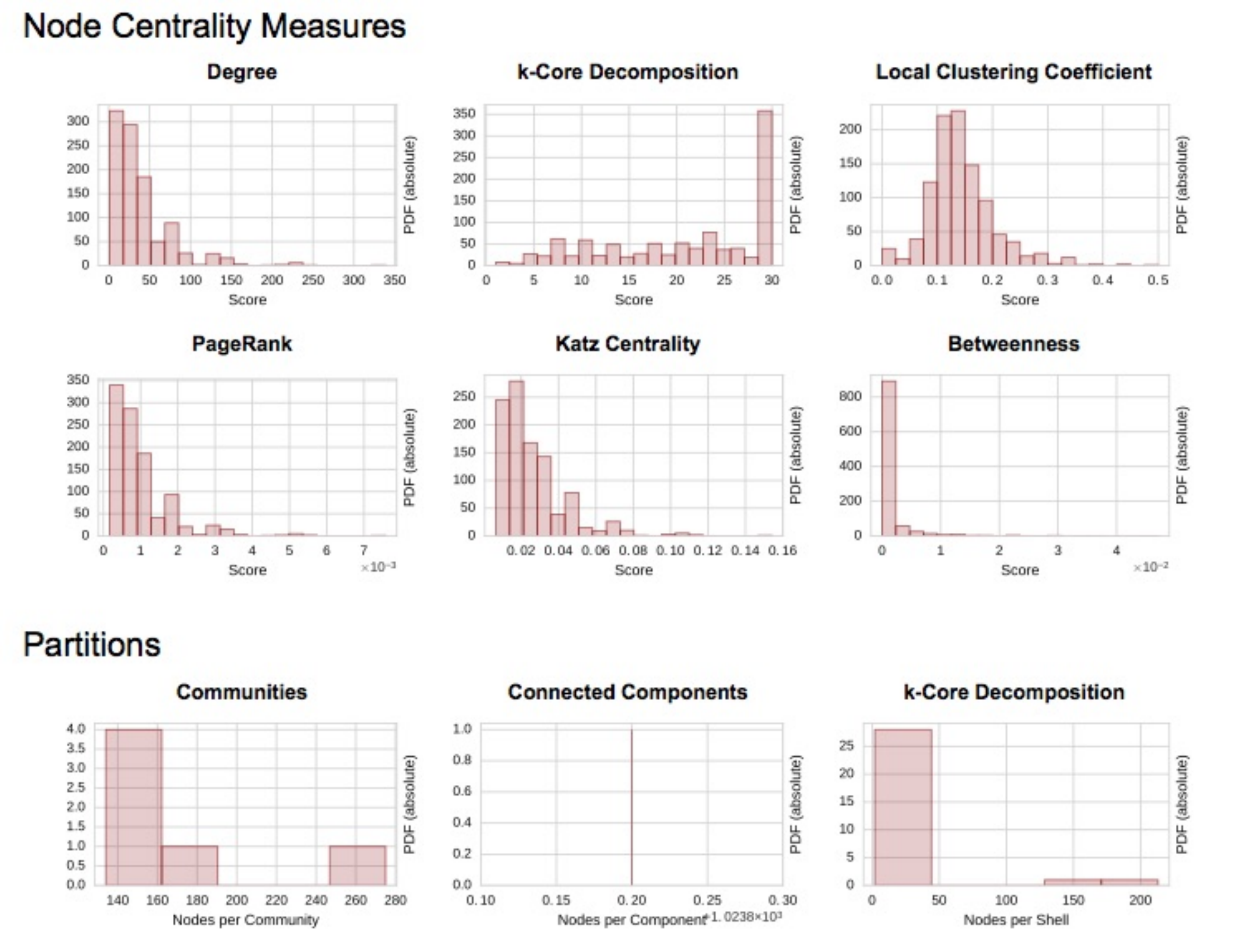}
}
\caption{(cont.) Structure profiles for the evaluation of the \textsf{kronfit} scheme}
\label{fig:kronfit-profiles2}
\end{center}
\end{figure}

\section{Appendix: Replicating Structural
Properties}\label{replicating-structural-properties}

In the existing literature on generative models, claims of realism are
typically substantiated by showing that a set of structural properties
is similar for real and synthetic networks. The large palette of
properties to choose from and the question which of those properties are
essential features makes this a complex problem. An often used approach
is to describe a network by a feature vector, a set of scalar
properties. These are often maxima, minima or averages of node
properties (cf.~the structural profiles published by the KONECT project
\cite{konect}). This can be reductionist, since these summary values
may not give enough information about how the node properties are
distributed. We therefore demonstrate how well the different models
replicate networks with two types of plots. The first plot type shows
scalar properties of the replicas, relative to those of the original.
Let \(x_{o}\) and \(x_{r}\) denote scalar properties of the original and
the replica, respectively. The relative value \(x_{r} / x_{o}\) is
computed for each replica. Box plots depict the distribution of these
relative values over the entire set of replicas. Scalar properties
included are: the number of edges, the number of connected components,
the effective diameter (for 90\% of node pairs) of the largest connected
component, and the number of communities. The effective diameter is
approximated using the \textsf{ANF} algorithm \cite{palmer2002anf}.
Communities are detected using the modularity-maximizing algorithm
\textsf{PLM}~\cite{DBLP:journals/tpds/StaudtM16}. This
set was chosen so that it could be quickly computed for a large set of
networks. The second type of plot covers centrality measures, and is
designed to show how the shape of the distributions of node centrality
scores of the originals compares to those of the replicas. Each segment
of the plot depicts the centrality values of all nodes of all networks
in the considered data set or the replicas of a certain algorithm. Since
centrality measures from this selection can have very different scales,
all centrality scores are normalized to the interval \([0,1]\).

\begin{figure}[p]
\begin{center}
  \subfigure[relative difference of scalar network properties]{
    \includegraphics[width=\textwidth]{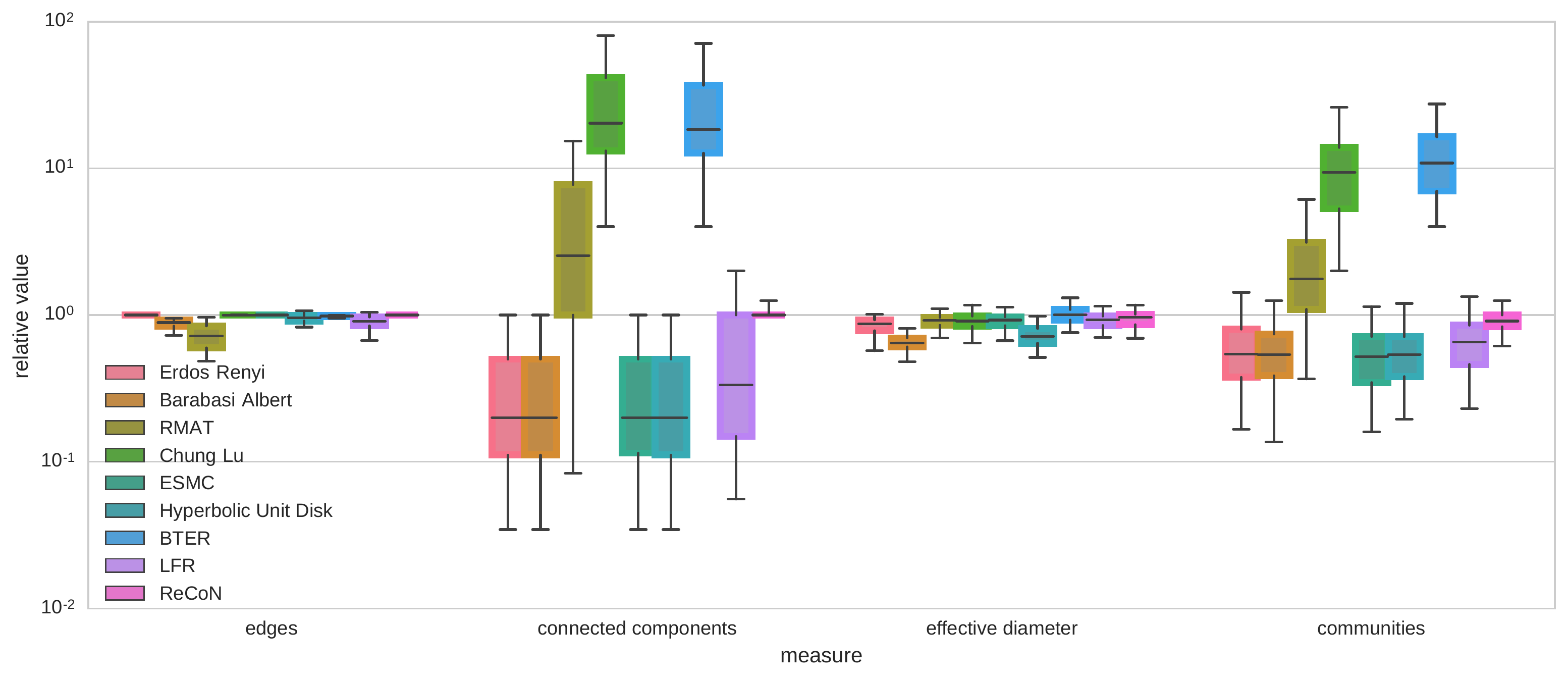}
    \label{fig:fb-structure-replication-scalars}
  }
\hfill
  \subfigure[distribution of centrality scores]{
    \includegraphics[width=\textwidth]{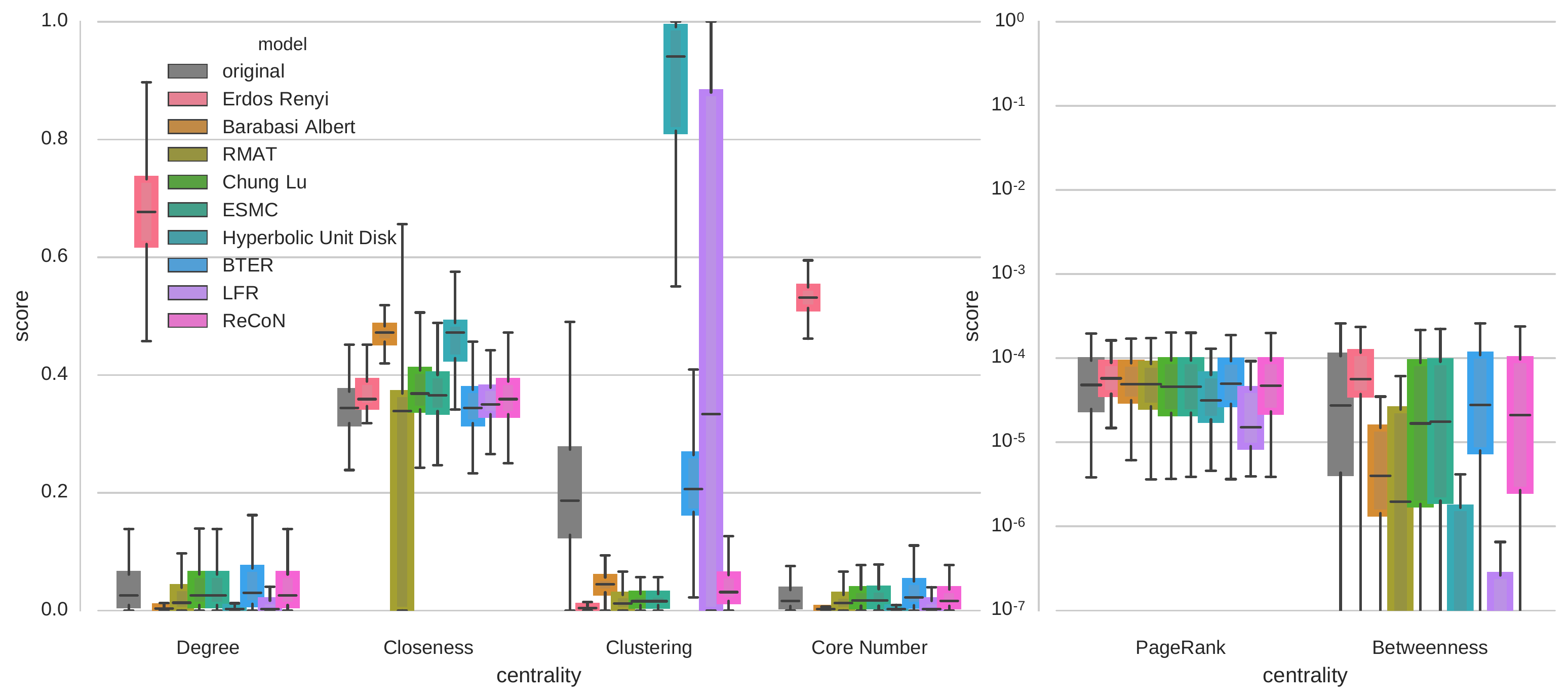}
    \label{fig:fb-structure-replication-centralities}
  }
\caption{Structure replication of Facebook networks}
\label{fig:fb-structure-replication}
\end{center}
\end{figure}

\begin{figure}[htb]
\begin{center}
  \subfigure[relative difference of scalar network properties]{
    \includegraphics[width=\textwidth]{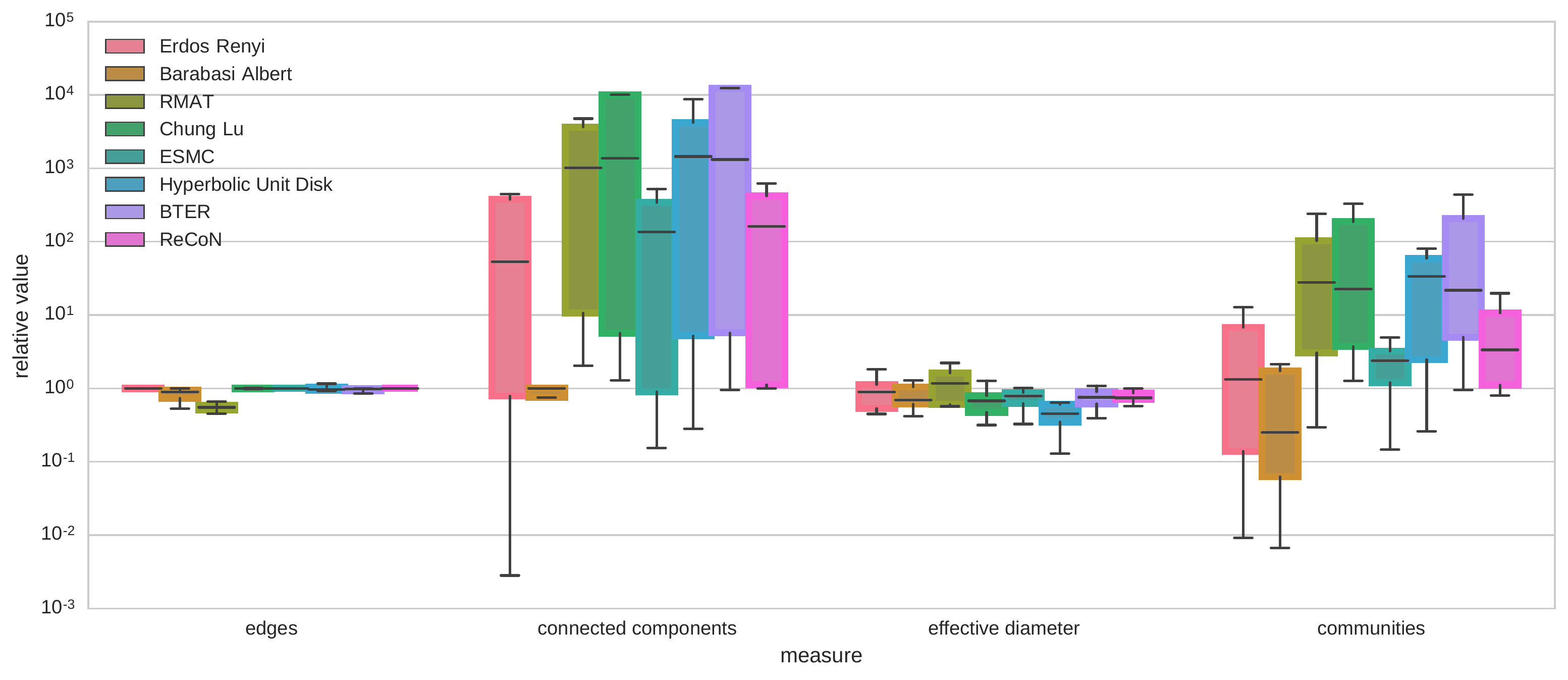}
    \label{fig:alt-structure-replication-scalars}
  }
\hfill
  \subfigure[distribution of centrality scores]{
    \includegraphics[width=\textwidth]{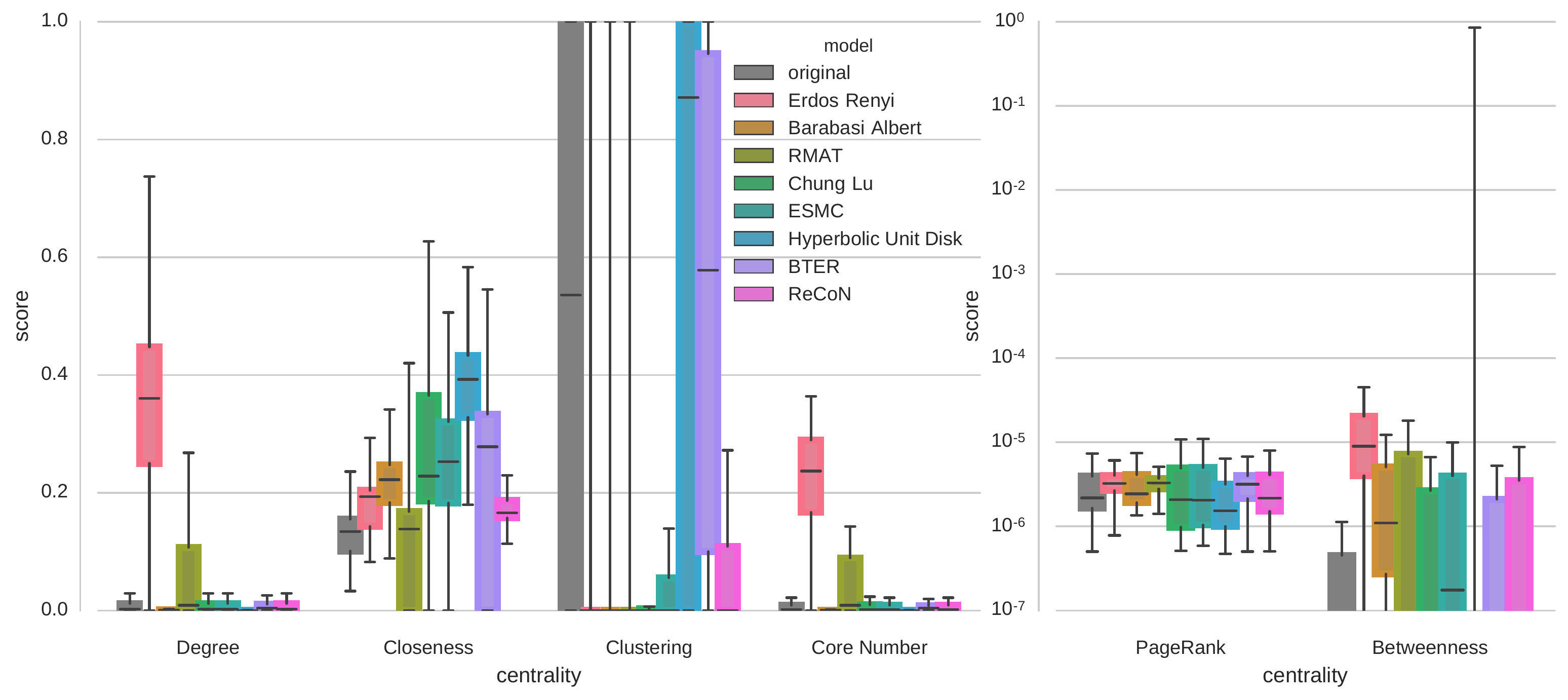}
    \label{fig:alt-structure-replication-centralities}
  }
\caption{Structure replication of diverse set of networks (Tab.~\ref{tab:add-networks})}
\label{fig:alt-structure-replication}
\end{center}
\end{figure}

Describing the results on 100 Facebook networks (Fig.
\ref{fig:fb-structure-replication-scalars}) from left to right, we
observe the following results for the measured scalar properties: All
models were parametrized to produce exactly \(n' = n\) nodes, while some
degree of freedom exists in the number of edges. However, all generators
replicate the number of edges with a narrow variance, which is largest
for RMAT. \recon is the only model that matches the number of components
with high accuracy, while it is lower for ER, BA, ESMC, HUDG and
LFR, and significantly higher for CL and BTER. The latter can probably
be explained by the generation of isolated nodes instead of degree-1
nodes which we have already seen in the visualizations before. There is
no extreme deviation in terms of the effective diameter, but note that
for small world networks, relatively small differences in the effective
diameter may indicate significant structural differences. The
Barabasi-Albert and Hyperbolic Unit Disk models deviate the most and
produce lower diameters. \recon, which receives a partition of the
original into communities as input, replicates it closely, but all yield
different numbers of communities. The Chung-Lu and BTER generators
increase the number of communities by a factor of 10 on average which
can be explained by the larger number of connected components. Overall,
\recon emerges with the most accurate replicas from this experiment.

The distributions of node centralities (Fig.
\ref{fig:fb-structure-replication-centralities}) compare as follows: All
models except ER are capable of producing skewed degree distributions,
with CL, ESMC, BTER and \recon matching the original closely. Closeness is
approximately matched by most models, but BA, RMAT and HUDG deviate
significantly. The original networks feature a wide range of clustering
coefficients, and only the BTER model, which receives explicitly this
distribution, matches them exactly. For HUDG, clustering is extremely
artificially high with a median close to 0.9, while LFR produces an
unrealistically large variance. For the \(k\)-core numbers, random
graphs are clearly outliers. RMAT, CL, ESMC, BTER and \recon match well,
while the very narrow distributions of the others point to a lack of
differentiated \(k\)-core structure. Only small variations exist with
respect to PageRank, and HUDG and LFR have strong deviations in
terms of betweenness centrality, but interpretation is not
straightforward in this case. In summary, extreme deviations in
centrality score distributions clearly give away the artificiality of
some synthetic graphs, such as the clustering coefficient for HUDG or
the degree distribution for ER. Other differences are more subtle, but
possibly relevant. BTER replicates centralities most accurately.

\begin{table}[htbp]
\begin{center}
\begin{small}
\begin{tabular}{llrr}
\bf Network               & \bf Type                         & \bf $n$  & \bf $m$ \\ \hline
email-Enron     & email communication   & 36691     &  183830 \\ \hline
PGPgiantcompo   & PGP web of trust  & 10680     &  24316 \\ \hline
as-22july06     & internet topology & 22963     &  48436 \\ \hline
hep-th  & scientific coauthorship   & 8361  &  15751 \\ \hline
dolphins    & animal social network & 62    &  159 \\ \hline
power   & power grid    & 4941  &  6594 \\ \hline
cnr-2000    & web graph & 325557    &  2738969 \\ \hline
\end{tabular}
\end{small}
\caption{Additional networks used. The network email-Enron has been taken from the Stanford Large Network Dataset Collection~\cite{snapnets}, all other networks are from the clustering instances of the 10th DIMACS implementation challenge~\cite{bader2014benchmarking}.}
\label{tab:add-networks}
\end{center}
\end{table}

The results on a different set of networks (Tab.~\ref{tab:add-networks}) are presented in Fig.
\ref{fig:alt-structure-replication}. A notable difference is the extreme
variance of clustering coefficients in the original set, which \recon
cannot replicate. Again \recon performs well for the majority of
properties.

\begin{figure}[htbp]
\begin{center}
\subfigure[relative difference of scalar network properties]{
    \includegraphics[width=\textwidth]{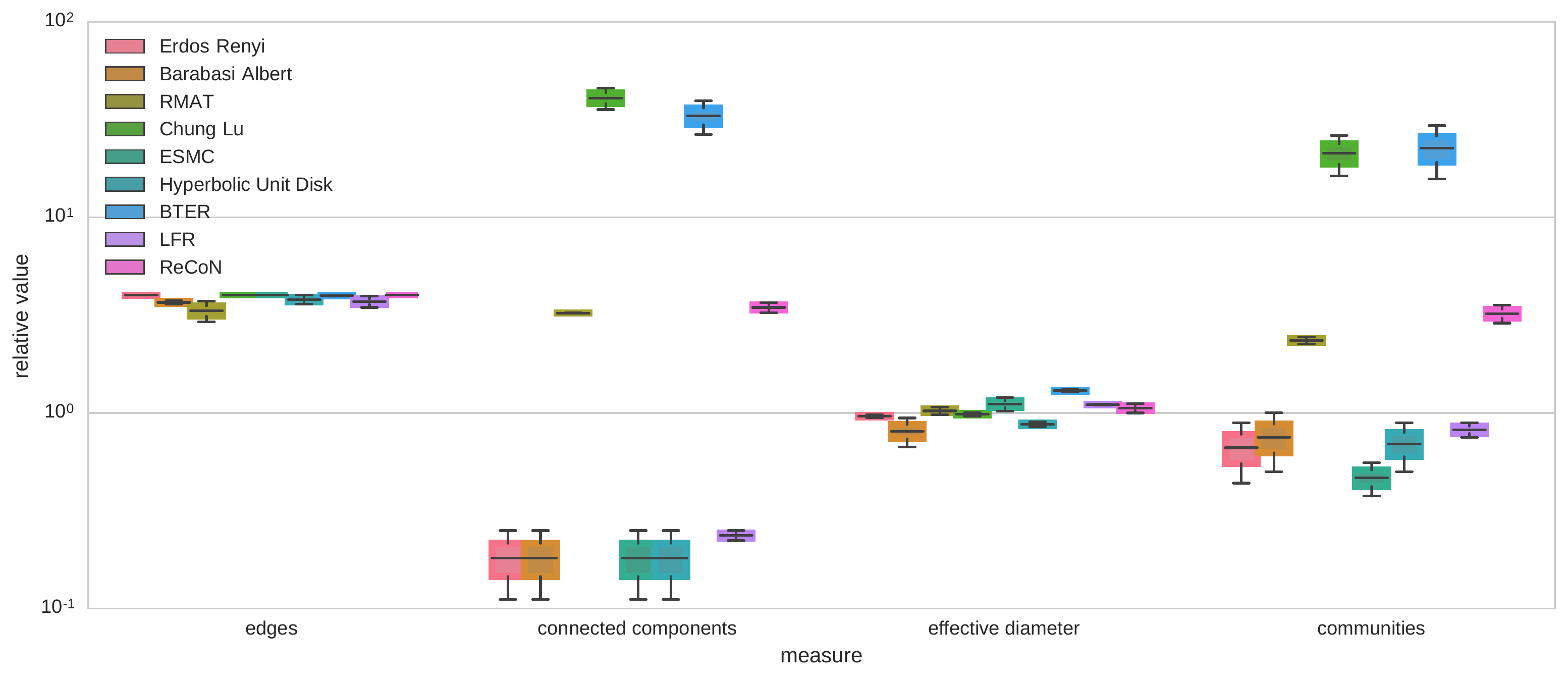}
    \label{fig:fb-structure-replication-scalars-scaled}
}
\hfill
\subfigure[distribution of centrality scores]{
    \includegraphics[width=\textwidth]{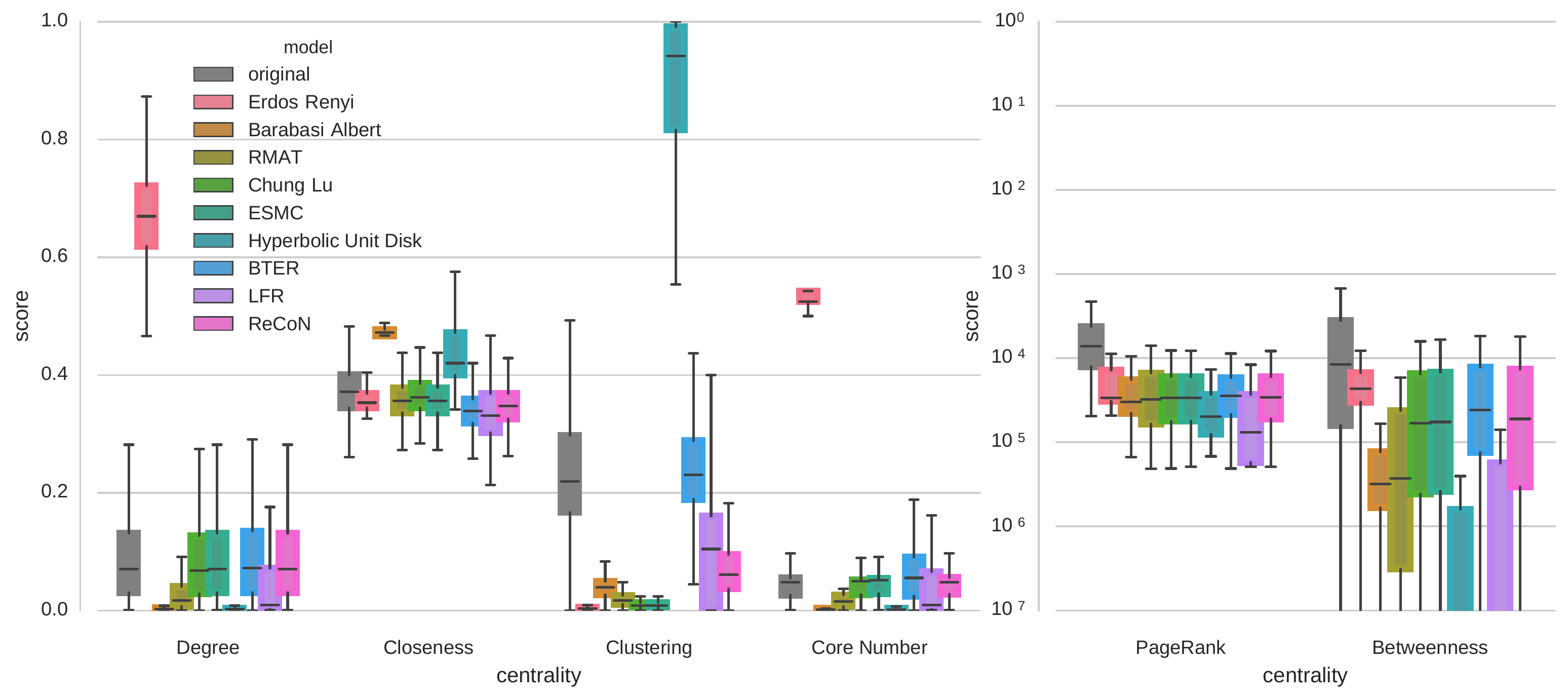}
    \label{fig:fb-structure-replication-centralities-scaled}
}
\caption{Structure replication of Facebook networks with scaling factor 4}
\label{fig:fb-structure-replication-scaled}
\end{center}
\end{figure}

Fig. \ref{fig:fb-structure-replication-scaled} shows results of a
repetition of the experiment with a scaling factor of 4. All models
except RMAT achieve the targeted edge factor of \(m' = 4 \cdot m\). The
number of components is unrealistically increased by RMAT and CL. For
the effective diameter, small relative differences matter: BA and HUDG
model tend to create smaller worlds than reality. \recon produces a
remarkably exact match, considering that the generator does not
explicitly target the diameter. It does however target a higher number
of communities, which is desired and achieved. LFR keeps the number of
communities constant on average, while many other models produce fewer
communities than the originals. The relative differences in the
distributions of centralities are qualitatively equivalent to those in
Fig. \ref{fig:fb-structure-replication-centralities}.


\begin{thebibliography}{10}

\bibitem{aiello2000random}
William Aiello, Fan Chung, and Linyuan Lu.
\newblock A random graph model for massive graphs.
\newblock In {\em Proceedings of the thirty-second annual ACM symposium on
  Theory of computing}, pages 171--180. Acm, 2000.

\bibitem{Albert2002}
R.~Albert and A.-L. Barab\'{a}si.
\newblock Statistical mechanics of complex networks.
\newblock {\em Reviews of Modern Physics}, 74(1):47--97, 2002.

\bibitem{albert2002statistical}
R.~Albert and A.L. Barab{\'a}si.
\newblock Statistical mechanics of complex networks.
\newblock {\em Reviews of modern physics}, 74(1):47, 2002.

\bibitem{bader2014benchmarking}
David~A Bader, Henning Meyerhenke, Peter Sanders, Christian Schulz, Andrea
  Kappes, and Dorothea Wagner.
\newblock Benchmarking for graph clustering and partitioning.
\newblock In {\em Encyclopedia of Social Network Analysis and Mining}, pages
  73--82. Springer, 2014.

\bibitem{Bansal09}
S.~Bansal, S.~Khandelwal, and L.A. Meyers.
\newblock Exploring biological network structure with clustered random
  networks.
\newblock {\em BMC Bioinformatics}, 10(1):405, 2009.

\bibitem{barabasi}
A.-L. Barab\'{a}si and R.~Albert.
\newblock Emergence of scaling in random networks.
\newblock {\em Science}, 286(5439):509--512, October 1999.

\bibitem{batagelj2005efficient}
Vladimir Batagelj and Ulrik Brandes.
\newblock Efficient generation of large random networks.
\newblock {\em Physical Review E}, 71(3):036113, 2005.

\bibitem{Binder08}
P.~M. Binder.
\newblock Frustration in complexity.
\newblock {\em Science}, 322:323, 2008.

\bibitem{boccaletti2006complex}
Stefano Boccaletti, Vito Latora, Yamir Moreno, Martin Chavez, and D-U Hwang.
\newblock Complex networks: Structure and dynamics.
\newblock {\em Physics reports}, 424(4):175--308, 2006.

\bibitem{brandt:review01}
A.~Brandt.
\newblock Multiscale scientific computation: Review 2001.
\newblock In T.~Barth, R.~Haimes, and T.~Chan, editors, {\em Multiscale and
  Multiresolution Methods (Proceedings of the Yosemite Educational Symposium,
  October 2000)}. Springer-Verlag, 2001.

\bibitem{brandt:optstrat}
A.~Brandt and D.~Ron.
\newblock Chapter 1 : Multigrid solvers and multilevel optimization strategies.
\newblock In J.~Cong and J.~R. Shinnerl, editors, {\em Multilevel Optimization
  and VLSICAD}. Kluwer, 2003.

\bibitem{brase2009modeling}
J.~M. Brase and D.~L. Brown.
\newblock Modeling, simulation and analysis of complex networked systems - a
  program plan.
\newblock {\em Lawrence Livermore National Laboratory}, 2009.

\bibitem{briggs}
W.~L. Briggs, V.~Henson, and S.~F. McCormick.
\newblock {\em A multigrid tutorial: second edition}.
\newblock Society for Industrial and Applied Mathematics, Philadelphia, PA,
  2000.

\bibitem{bulucc2016recent}
Ayd{\i}n Bulu{\c{c}}, Henning Meyerhenke, Ilya Safro, Peter Sanders, and
  Christian Schulz.
\newblock Recent advances in graph partitioning.
\newblock In {\em Algorithm Engineering}, pages 117--158. Springer, 2016.

\bibitem{caldarelli2007large}
Guido Caldarelli and Alessandro Vespignani.
\newblock {\em Large scale structure and dynamics of complex networks}.
\newblock World Scientific, 2007.

\bibitem{Carlson02}
J.M. Carlson and J.~Doyle.
\newblock Complexity and robustness.
\newblock {\em Proceedings of the National Academy of Sciences of the United
  States of America}, 99(Suppl 1):2538, 2002.

\bibitem{chakrabarti2006graph}
Deepayan Chakrabarti and Christos Faloutsos.
\newblock Graph mining: Laws, generators, and algorithms.
\newblock {\em ACM Computing Surveys (CSUR)}, 38(1):2, 2006.

\bibitem{chakrabarti2004r}
Deepayan Chakrabarti, Yiping Zhan, and Christos Faloutsos.
\newblock {R-MAT: A recursive model for graph mining}.
\newblock {\em Computer Science Department}, page 541, 2004.

\bibitem{clauset2009power}
Aaron Clauset, Cosma~Rohilla Shalizi, and Mark~EJ Newman.
\newblock Power-law distributions in empirical data.
\newblock {\em SIAM review}, 51(4):661--703, 2009.

\bibitem{DBLP:conf/bigdataconf/DasariRZ14}
Naga~Shailaja Dasari, Desh Ranjan, and Mohammad Zubair.
\newblock {ParK: An efficient algorithm for k-core decomposition on multicore
  processors}.
\newblock In {\em 2014 {IEEE} International Conference on Big Data, Big Data
  2014, Washington, DC, USA, October 27-30, 2014}, pages 9--16. {IEEE}, 2014.

\bibitem{dunlavy2009mathematical}
D.M. Dunlavy, B.~Hendrickson, and T.G. Kolda.
\newblock Mathematical challenges in cybersecurity.
\newblock Technical report, Sandia National Laboratory, 2009.

\bibitem{erdos1960erg}
P.~Erdos and A.~Renyi.
\newblock On the evolution of random graphs.
\newblock {\em Publ. Math. Inst. Hung. Acad. Sci}, 5:17--61, 1960.

\bibitem{Eubank04}
S.~Eubank, H.~Guclu, V.S.A. Kumar, M.V. Marathe, A.~Srinivasan, Z.~Toroczkai,
  and N.~Wang.
\newblock Modelling disease outbreaks in realistic urban social networks.
\newblock {\em Nature}, 429(6988):180--184, 2004.

\bibitem{lfrimpl}
Santo Fortunato.
\newblock Benchmark graphs to test community detection algorithms.
\newblock \url{https://sites.google.com/site/santofortunato/inthepress2}.

\bibitem{geisberger2008better}
Robert Geisberger, Peter Sanders, and Dominik Schultes.
\newblock Better approximation of betweenness centrality.
\newblock In {\em ALENEX}, pages 90--100. SIAM, 2008.

\bibitem{glantz2016tree}
Roland Glantz, Henning Meyerhenke, and Christian Schulz.
\newblock Tree-based coarsening and partitioning of complex networks.
\newblock {\em Journal of Experimental Algorithmics (JEA)}, 21(1):1--6, 2016.

\bibitem{goldenberg2010survey}
Anna Goldenberg, Alice~X Zheng, Stephen~E Fienberg, and Edoardo~M Airoldi.
\newblock A survey of statistical network models.
\newblock {\em Foundations and Trends{\textregistered} in Machine Learning},
  2(2):129--233, 2010.

\bibitem{musketeer-impl}
A.~Gutfraind, L.A. Meyers, and I.~Safro.
\newblock {MUSKETEER}: Multiscale entropic network generator.
\newblock \url{https://people.cs.clemson.edu/~isafro/musketeer/index.html},
  2012.

\bibitem{gutfraind2015multiscale}
Alexander Gutfraind, Ilya Safro, and Lauren~Ancel Meyers.
\newblock Multiscale network generation.
\newblock In {\em Information Fusion (Fusion), 2015 18th International
  Conference on}, pages 158--165. IEEE, 2015.

\bibitem{DBLP:journals/snam/HamannLMSW16}
Michael Hamann, Gerd Lindner, Henning Meyerhenke, Christian~L. Staudt, and
  Dorothea Wagner.
\newblock Structure-preserving sparsification methods for social networks.
\newblock {\em Social Netw. Analys. Mining}, 6(1):22:1--22:22, 2016.

\bibitem{Itzkovitz05}
S.~Itzkovitz, R.~Levitt, N.~Kashtan, R.~Milo, M.~Itzkovitz, and U.~Alon.
\newblock Coarse-graining and self-dissimilarity of complex networks.
\newblock {\em Physical Review E}, 71(1):016127, 2005.

\bibitem{Keeling_Rohani_2008}
Matthew~James Keeling and Pejman Rohani.
\newblock {\em Modeling infectious diseases in humans and animals}, volume~47.
\newblock Princeton University Press, 2008.

\bibitem{kolda2013scalable}
Tamara~G Kolda, Ali Pinar, Todd Plantenga, and C~Seshadhri.
\newblock A scalable generative graph model with community structure.
\newblock {\em arXiv preprint arXiv:1302.6636}, 2013.

\bibitem{krapivsky}
P.~L. Krapivsky and S.~Redner.
\newblock Organization of growing random networks.
\newblock {\em Phys. Rev. E}, 63(066123), 2001.

\bibitem{Krioukov2010}
Dmitri Krioukov, Fragkiskos Papadopoulos, Maksim Kitsak, Amin Vahdat, and
  Mari\'an Bogu\~n\'a.
\newblock Hyperbolic geometry of complex networks.
\newblock {\em Physical Review E}, 82:036106, Sep 2010.

\bibitem{konect}
J{\'e}r{\^o}me Kunegis.
\newblock Konect: the koblenz network collection.
\newblock In {\em Proceedings of the 22nd international conference on World
  Wide Web companion}, pages 1343--1350. International World Wide Web
  Conferences Steering Committee, 2013.

\bibitem{lancichinetti2009benchmarks}
Andrea Lancichinetti and Santo Fortunato.
\newblock Benchmarks for testing community detection algorithms on directed and
  weighted graphs with overlapping communities.
\newblock {\em Physical Review E}, 80(1):016118, 2009.

\bibitem{lancichinetti2008benchmark}
Andrea Lancichinetti, Santo Fortunato, and Filippo Radicchi.
\newblock Benchmark graphs for testing community detection algorithms.
\newblock {\em Physical Review E}, 78(4):046110, 2008.

\bibitem{kronmodel}
J.~Leskovec, D.~Chakrabarti, J.~Kleinberg, C.~Faloutsos, and Z.~Ghahramani.
\newblock Kronecker graphs: An approach to modeling networks.
\newblock {\em The Journal of Machine Learning Research}, 11:985--1042, 2010.

\bibitem{Leskovec08}
J.~Leskovec, K.J. Lang, A.~Dasgupta, and M.W. Mahoney.
\newblock Statistical properties of community structure in large social and
  information networks.
\newblock In {\em Proceeding of the 17th international conference on World Wide
  Web}, pages 695--704. ACM, 2008.

\bibitem{leskovec2007scalable}
Jure Leskovec and Christos Faloutsos.
\newblock Scalable modeling of real graphs using kronecker multiplication.
\newblock In {\em Proc. 24th Intl. Conference on Machine learning}, pages
  497--504. ACM, 2007.

\bibitem{snapnets}
Jure Leskovec and Andrej Krevl.
\newblock {SNAP Datasets}: {Stanford} large network dataset collection.
\newblock \url{http://snap.stanford.edu/data}, June 2014.

\bibitem{snap}
Jure Leskovec and Rok Sosi\v{c}.
\newblock {SNAP}: A general purpose network analysis and graph mining library
  in {C++}.
\newblock \url{http://snap.stanford.edu/snap}, June 2014.

\bibitem{leyffer2013fast}
Sven Leyffer and Ilya Safro.
\newblock Fast response to infection spread and cyber attacks on large-scale
  networks.
\newblock {\em Journal of Complex Networks}, 1(2):183--199, 2013.

\bibitem{mahadevan2006systematic}
P.~Mahadevan, D.~Krioukov, K.~Fall, and A.~Vahdat.
\newblock Systematic topology analysis and generation using degree
  correlations.
\newblock {\em ACM SIGCOMM Computer Communication Review}, 36(4):135--146,
  2006.

\bibitem{medina}
Alberto Medina, Ibrahim Matta, and John Byers.
\newblock On the origin of power laws in internet topologies.
\newblock Technical report, Boston University, Boston, MA, USA, 2000.

\bibitem{DBLP:conf/gd/MeyerhenkeN015}
Henning Meyerhenke, Martin N{\"{o}}llenburg, and Christian Schulz.
\newblock Drawing large graphs by multilevel maxent-stress optimization.
\newblock In Emilio~Di Giacomo and Anna Lubiw, editors, {\em Graph Drawing and
  Network Visualization - 23rd International Symposium, {GD} 2015, Los Angeles,
  CA, USA, September 24-26, 2015, Revised Selected Papers}, volume 9411 of {\em
  Lecture Notes in Computer Science}, pages 30--43. Springer, 2015.

\bibitem{DBLP:journals/heuristics/MeyerhenkeS016}
Henning Meyerhenke, Peter Sanders, and Christian Schulz.
\newblock Partitioning (hierarchically clustered) complex networks via
  size-constrained graph clustering.
\newblock {\em J. Heuristics}, 22(5):759--782, 2016.

\bibitem{Meyers05}
L.A. Meyers, B.~Pourbohloul, M.E.J. Newman, D.M. Skowronski, and R.C. Brunham.
\newblock Network theory and sars: predicting outbreak diversity.
\newblock {\em Journal of theoretical biology}, 232(1):71--81, 2005.

\bibitem{mihail2003markov}
C.G.M. Mihail and E.~Zegura.
\newblock The markov chain simulation method for generating connected power law
  random graphs.
\newblock In {\em Proceedings of the Fifth Workshop on Algorithm Engineering
  and Experiments}, volume 111, page~16. Society for Industrial Mathematics,
  2003.

\bibitem{milo:2003}
R.~{Milo}, N.~{Kashtan}, S.~{Itzkovitz}, M.~E.~J. {Newman}, and U.~{Alon}.
\newblock {On the uniform generation of random graphs with prescribed degree
  sequences}.
\newblock {\em eprint arXiv:cond-mat/0312028}, December 2003.

\bibitem{Mones12}
Enys Mones, Lilla Vicsek, and Tamas Vicsek.
\newblock Hierarchy measure for complex networks.
\newblock {\em PLoS ONE}, 7(3):e33799, 03 2012.

\bibitem{Newman03thestructure}
M.~E.~J. Newman.
\newblock The structure and function of complex networks.
\newblock {\em SIAM Review}, 45:167--256, 2003.

\bibitem{Newman2010networks}
Mark Newman.
\newblock {\em Networks: an introduction}.
\newblock Oxford University Press, 2010.

\bibitem{palla2010multifractal}
G.~Palla, L.~Lov{\'a}sz, and T.~Vicsek.
\newblock Multifractal network generator.
\newblock {\em Proceedings of the National Academy of Sciences}, 107(17):7640,
  2010.

\bibitem{palla2005uncovering}
Gergely Palla, Imre Der{\'e}nyi, Ill{\'e}s Farkas, and Tam{\'a}s Vicsek.
\newblock Uncovering the overlapping community structure of complex networks in
  nature and society.
\newblock {\em Nature}, 435(7043):814--818, 2005.

\bibitem{palmer2002anf}
Christopher~R Palmer, Phillip~B Gibbons, and Christos Faloutsos.
\newblock Anf: A fast and scalable tool for data mining in massive graphs.
\newblock In {\em Proceedings of the eighth ACM SIGKDD international conference
  on Knowledge discovery and data mining}, pages 81--90. ACM, 2002.

\bibitem{potterat2002risk}
John~J Potterat, L~Phillips-Plummer, Stephen~Q Muth, RB~Rothenberg,
  DE~Woodhouse, TS~Maldonado-Long, HP~Zimmerman, and JB~Muth.
\newblock Risk network structure in the early epidemic phase of hiv
  transmission in colorado springs.
\newblock {\em Sexually transmitted infections}, 78(suppl 1):i159--i163, 2002.

\bibitem{safro:relaxml}
D.~Ron, I.~Safro, and A.~Brandt.
\newblock Relaxation-based coarsening and multiscale graph organization.
\newblock {\em Multiscale Modeling {\&} Simulation}, 9(1):407--423, 2011.

\bibitem{schlauch:2015}
Wolfgang~E. Schlauch, Em{\H{o}}ke~{\'A}gnes Horv{\'a}t, and Katharina~A. Zweig.
\newblock Different flavors of randomness: comparing random graph models with
  fixed degree sequences.
\newblock {\em Social Network Analysis and Mining}, 5(1):1--14, 2015.

\bibitem{Seshadhri2011}
C.~Seshadhri, Tamara~G. Kolda, and Ali Pinar.
\newblock {Community structure and scale-free collections of Erd{\"o}s-Renyi
  graphs}.
\newblock December 2011.

\bibitem{snijders2001statistical}
Tom~AB Snijders.
\newblock The statistical evaluation of social network dynamics.
\newblock {\em Sociological methodology}, 31(1):361--395, 2001.

\bibitem{staudt16algorithms}
Christian~L. Staudt.
\newblock {\em Algorithms and Software for the Analysis of Large Complex
  Networks}.
\newblock PhD thesis, Karlsruhe Institute of Technology, June 2016.

\bibitem{staudt2016generating}
Christian~L Staudt, Michael Hamann, Ilya Safro, Alexander Gutfraind, and
  Henning Meyerhenke.
\newblock Generating scaled replicas of real-world complex networks.
\newblock In {\em International Workshop on Complex Networks and their
  Applications}, pages 17--28. Springer, 2016.

\bibitem{DBLP:journals/tpds/StaudtM16}
Christian~L. Staudt and Henning Meyerhenke.
\newblock Engineering parallel algorithms for community detection in massive
  networks.
\newblock {\em {IEEE} Trans. on Parallel and Distributed Systems},
  27(1):171--184, 2016.

\bibitem{StaudtSM16networkit}
Christian~L. Staudt, Aleksejs Sazonovs, and Henning Meyerhenke.
\newblock {NetworKit}: A tool suite for large-scale complex network analysis.
\newblock {\em Network Science}, 4(4):508–530, 2016.

\bibitem{feastpack}
Sandia National~Laboratories Tamara G.~Kolda, Ali~Pinar.
\newblock Feastpack.
\newblock \url{http://www.sandia.gov/~tgkolda/feastpack/}.

\bibitem{traud2012social}
Amanda~L Traud, Peter~J Mucha, and Mason~A Porter.
\newblock Social structure of facebook networks.
\newblock {\em Physica A: Statistical Mechanics and its Applications},
  391(16):4165--4180, 2012.

\bibitem{mgbooktrott}
U.~Trottenberg and A.~Schuller.
\newblock {\em Multigrid}.
\newblock Academic Press, Inc., 2001.

\bibitem{viger2005random}
Fabien Viger and Matthieu Latapy.
\newblock Random generation of large connected simple graphs with prescribed
  degree distribution.
\newblock In {\em 11th International Conference on Computing and Combinatorics.
  Kunming, Yunnan, Chine}, 2005.

\bibitem{DBLP:conf/isaac/LoozMP15}
Moritz von Looz, Henning Meyerhenke, and Roman Prutkin.
\newblock Generating random hyperbolic graphs in subquadratic time.
\newblock In {\em Algorithms and Computation - 26th International Symposium,
  {ISAAC} 2015, Nagoya, Japan, December 9-11, 2015, Proceedings}, pages
  467--478, 2015.

\bibitem{Watts1998}
D~J Watts and S~H Strogatz.
\newblock {Collective dynamics of 'small-world' networks.}
\newblock {\em Nature}, 393(6684):440--2, June 1998.

\bibitem{Wolpert07}
D.H. Wolpert and W.~Macready.
\newblock Using self-dissimilarity to quantify complexity.
\newblock {\em Complexity}, 12(3):77--85, 2007.

\end{thebibliography}
\end{document}